\documentclass[aps,prl,reprint]{revtex4-1}

\usepackage{graphicx}

\usepackage{color}\usepackage{enumerate}
\usepackage{amsmath}\usepackage{amssymb}\usepackage{stmaryrd}
\usepackage{multirow}
\usepackage{float}
\usepackage{longtable,booktabs}
\usepackage{color}\usepackage{enumerate}
\usepackage{amsmath}\usepackage{amssymb}\usepackage{stmaryrd}
\usepackage{multirow}
\usepackage{float}
\usepackage[toc,page]{appendix}
\usepackage{longtable,booktabs}
 \usepackage{epsfig}
\usepackage{graphicx} 
\usepackage{feynmf}
\usepackage{psfrag} 
\usepackage{slashed}
\usepackage{amsmath,amssymb} 
\usepackage{colordvi} 
\usepackage{hyperref}
\usepackage{setspace}
\usepackage[all]{hypcap}
\usepackage{natbib}
\usepackage{subfigure}

\hypersetup{
    bookmarks=true,         % show bookmarks bar?
    unicode=false,          % non-Latin characters in AcrobatÕs bookmarks
    pdftoolbar=true,        % show AcrobatÕs toolbar?
    pdfmenubar=true,        % show AcrobatÕs menu?
    pdffitwindow=false,     % window fit to page when opened
    pdfstartview={FitH},    % fits the width of the page to the window
    pdftitle={My title},    % title
    pdfauthor={Author},     % author
    pdfsubject={Subject},   % subject of the document
    pdfcreator={Creator},   % creator of the document
    pdfproducer={Producer}, % producer of the document
    pdfkeywords={keyword1} {key2} {key3}, % list of keywords
    pdfnewwindow=true,      % links in new window
    colorlinks=true,       % false: boxed links; true: colored links
    linkcolor=red,          % color of internal links
    citecolor=green,        % color of links to bibliography
    filecolor=magenta,      % color of file links
    urlcolor=black           % color of external links
}
\def\be{\begin{equation}}
\def\bea{\begin{eqnarray}}
\def\eea{\end{eqnarray}}
\def\ee{\end{equation}}
\def\del{\partial}

\begin{document}

\title{Quasi-Topological Electromagnetic Response of Line-node Semimetals}
\author{Srinidhi T. Ramamurthy}
\author{Taylor L. Hughes}
\affiliation{Department of Physics, Institute for Condensed Matter Theory, University of Illinois at Urbana-Champaign, IL 61801, USA}

\begin{abstract}
Topological semimetals are gapless states of matter which have robust surface states and interesting electromagnetic responses. In this paper, we consider the electromagnetic response of gapless phases in $3+1$-dimensions with line nodes. We show through a layering approach that an intrinsic $2$-form ${\cal{B}}_{\mu\nu}$ emerges in the effective response field theory that is determined by the geometry and energy-embedding of the nodal lines. This 2-form is shown to be simply related to the charge polarization and orbital magnetization of the sample. We conclude by discussing the relevance for recently proposed materials and heterostructures with line-node fermi-surfaces. 
\end{abstract}

%\pacs{}
\maketitle
%\tableofcontents
Topological insulators (TIs) have been of great interest in recent years after their theoretical proposal and experimental discovery in the past decade. Their electronic properties led to a wide search for novel topological band structures in many materials\cite{HasanKane,bernevigbook}. TIs are characterized by a gapped bulk and protected boundary modes that are robust in the presence of disorder. They also exhibit quantized properties in their electromagnetic (EM) response\cite{haldane1988,qi2008}. A classification of non-interacting fermionic states protected by discrete time-reversal ($\mathcal{T}$), charge-conjugation ($\mathcal{C}$), and chiral symmetries has been worked out in Refs. \onlinecite{schnyder2008,qi2008,kitaev2009}. This has further been expanded on in recent years to include translation, reflection, and rotation symmetries of crystalline systems\cite{FuKaneWeak2007,teo2008,fu2011topological,hughes2011inversion,turner2012, fang2012bulk,teohughes1,slager2012, Benalcazar2013,Morimoto2013,RyuReflection2013,fang2013entanglement,HughesYaoQi13,Sato2013,Kane2013Mirror,jadaun2013}. These theoretical advances have been accompanied by experimental discoveries of several TIs in various symmetry classes. The 3D $\mathcal{T}$-invariant strong TI (e.g., BiSb\cite{hsieh08}, Bi$_2$Se$_3$\cite{moore2009topological,xia2009observation,zhang2009experimental}), the 2D quantum spin Hall insulator (e.g., CdTe/HgTe quantum wells\cite{Kane2005A,bernevig2006c,konig2008}), the 2D quantum anomalous Hall (Chern) insulator (e.g., Cr-doped (Bi,Sb)$_2$Te$_3$\cite{haldane1988,chang2013}), and a 3D topological crystalline insulator (PbSnTe)\cite{xu2012observation,tanakatci}.

A defining characteristic of TIs is a gapped bulk, but one can also ask if there are gapless states of matter which harbor protected boundary modes and have unusual EM responses and transport properties. This question has been asked, and answered in the affirmative with the discovery of topological semimetals (TSMs). The most studied TSMs all have point-like Fermi surfaces, e.g.,  2D Dirac semi-metals/graphene\cite{graphenereview}, 3D Weyl semimetals\cite{wan2011,balentshalasz}, and 3D Dirac semimetals\cite{youngteo2012,wang2012,liu2013,neupane2013,wang2013,yang2014}. %Superconducting relatives of these have been proposed in \cite{meng2012,cho2012,matsuura2013}. 
In recent work, we proposed a unifying structure to understand TSMs with point-like Fermi-surfaces, from which one can straightforwardly determine the quasi-topological EM responses\cite{TSMresponse}, and which expands on previous work\cite{nielsen1981,niu2007,niuvalley,zyuzin2012,vazifeh2013,chen2013,haldane2014}. The main perspective which helps us understand these TSMs are models produced by a layering construction.  Generically, a TI phase in $d$ spatial dimensions can be layered/stacked into $d+1$ dimensions by introducing ``trivial" tunneling between the layers, i.e., tunneling that does not immediately generate a $d+1$-dimensional \emph{strong} topological phase. As the tunneling coefficient is increased, we showed that generically the material will transition from a weak topological insulator phase, which is formed in the decoupled limit, to a trivial insulating state with an intervening semimetallic gapless \emph{phase} with point-nodes. To ensure the stability of the gapless phase additional symmetries are often required. For example, in the case of the 2D Dirac semimetal, the stability of the intermediate TSM phase relies on the presence of a composite  spatial and anti-unitary symmetry, i.e., $\mathcal{T}\mathcal{I}$  where $\mathcal{I}$ is inversion symmetry.

The EM response of point-node TSMs is generally characterized by an intrinsic $1$-form $b=(b_{0},b_i)$ which is related to the locations of the nodes in momentum and energy space\cite{zyuzin2012,TSMresponse,niuvalley,niu2007}. This  quantity is  analogous to the weak invariant $\vec{\nu}=\tfrac{1}{2}\sum\nu_{i}\vec{G_{i}}$ (where $\nu_i$ are integers, and $\vec{G_{i}}$ are reciprocal lattice vectors)  that exists in the related, fully-gapped weak TI phase, but it additionally contains an extra time component $b_{0}$ due to the energy difference between the nodes. The time-component is ill-defined in the gapped phase unless, e.g., the system is subjected to a periodic driving frequency. The actual dependence of the EM response on $b$ depends on the type of point-node semimetal, and can generate a wide variety of effects in 2D and 3D TSMs. 

While a TSM with point-nodes arises from coupling $d$-dimensional topological phases into a $d+1$-dimensional system,
we can extend this idea by layering a $d$-dimensional topological phase into $d+2$ dimensions. When the $d$-dimensional elements are decoupled, the $d+2$-dimensional system will be in a secondary weak topological phase characterized by an anti-symmetric tensor/$2$-form invariant $\nu_{ij}$\cite{ran2010,hughesyaoqi}. When the lower dimensional topological phases are coupled with strong-enough ``trivial" hopping then they will produce line-node Fermi surfaces (FLs). We mentioned in Ref. \onlinecite{TSMresponse} that in these gapless phases we expect the EM response to be characterized by an analogous $2$-form $B_{\mu\nu}$. Hence, the goal of this article is to conclusively show that the effective response action is given by 
\be
\label{eq:action}
S[A,{\cal{B}}]=\frac{e}{16\pi^2}\int d^{4}x\,\epsilon^{\mu\nu\rho\sigma}{\cal{B}}_{\mu\nu}F_{\rho\sigma}
\ee\noindent where $B_{\mu\nu}$ is related to the magnetization and polarization via $e {\cal{B}}_{0i}=4\pi^2 M_{i}$ and $e {\cal{B}}_{ij}=4\pi^2 \epsilon_{ijk}P^{k}$ for $i=x,y,z.$  The $2$-form ${\cal{B}}_{\mu\nu}$ is an intrinsic property of line-node semimetals determined by the geometry of the nodal submanifolds, and is the analogue of a secondary weak invariant, though for a gapless phase. It  also includes components where $\mu,\nu$ are in the time direction, which are not available for a time-independent gapped system. 

To summarize, in this article, we consider the EM response of 3D TSMs with non-degenerate line-like Fermi surfaces which we will term  ``LTSMs." We note that a gapped version of this stacking construction has also been considered in \cite{yakovenko}, while a superconducting version of this, including line nodes, has been considered in \cite{chang2014,matsuura2013}. 
Additionally, in very recent work, several proposals for materials that realize line-node TSM states have appeared which utilize magnetic heterostructures\cite{burkov2011,phillips2014}, carbon allotropes \cite{uchoa2014,weng2014}, and inversion symmetric $\mathrm{Cu_{3}PdN}$\cite{Kane2015,rui2015}. After our primary discussion, we comment on how our analysis could be used to generate an EM response in these systems, including systems with nominally spin-degenerate FLs. 

%\begin{figure}
%\label{fig:fs}
%   \includegraphics[width=11cm]{fermistring.pdf}
%   \caption{A schematic illustrating the half Brillouin Zone in the case of the model Eq.~\ref{eq:fermistring} is shown in the figure. The colored regions show the topological regions enclosed by the Fermi strings which have edge states. These regions contribute to the bulk Polarization in the $x$-direction. The boundary of the colored regions is gapless and is the Fermi surface of the model.}
%\end{figure}

To aid our discussion it will be helpful to consider an explicit model. Let us take the 3D Bloch Hamiltonian
\bea
\label{eq:fermistring}
H_{3}(k)&=& \sin k_x\sigma^y + (1+\beta+\gamma-m-\cos k_x-\\\nonumber
&&\beta \cos k_y-\gamma\cos k_z)\sigma^z,
\eea\noindent  which has inversion $\mathcal{I}=\sigma^{z}$ and time reversal $\mathcal{T}=\sigma^{z}K$ symmetries, where $\sigma^a$ represent two (non-spin) degrees of freedom, and the lattice constant $a=1.$ When $\beta=\gamma=0$, this model reduces to decoupled 1D insulators aligned parallel to the $x$-direction. Since each 1D wire is inversion symmetric, their polarization will be quantized (and all equal). In their topological phase, the polarization of a single wire will be $P_{x}(k_y,k_z)=\frac{e}{2\pi}\int\mbox{Tr}[\,\mathcal{A}_{x}(k)]dk_x=e/2\,\mbox{mod}\, e$\cite{zak1989berry,vanderbilt1993,qhz2008,hughesprodanbernevig,turner2012},  where $\mathcal{A}_{i}(k_{x})$ is the adiabatic connection matrix $\mathcal{A}^{ab}_{i}(k)=-i\langle u_{a,k}\vert \tfrac{d}{dk_{i}}\vert u_{b,k}\rangle,$ where $a,b$ run over the occupied bands. If each insulator was instead in a trivial state, we would have $P_{x}(k_y,k_z)=0\,\mbox{mod}\,e$. The total polarization is just the sum over all the decoupled wires. In addition to the bulk topological properties, the 1D TIs have degenerate mid-gap modes localized at opposite ends of the system, the filling of which  determines the bound surface charge. To unambiguously determine the sign of the bulk polarization, and hence the sign of the surface charge, one must break the degeneracy by adding an infinitesimal (inversion) symmetry breaking mass, e.g.,  $m_{{\cal{I}}}\sigma^{y}$ and take the limit as $m_{{\cal{I}}}\to 0.$ Hence, $\beta=\gamma=0$ implies a secondary weak TI state protected by inversion symmetry and the EM response is given by Eq. \ref{eq:action}, but for the special case when ${\cal{B}}=\mbox{sgn}\,m_{{\cal{I}}}(\tfrac{1}{2}{\bf{G}}_{y}\wedge {\bf{G}}_{z})\implies P_x=\mbox{sgn}\,m_{{\cal{I}}}\tfrac{e}{2a_y a_z}.$ 

Now let us tune away from the decoupled limit. When $\gamma=0$ and $\beta$ is increased such that the bulk gap closes, we get layers of 2D Dirac semimetals with two gapless points in the Brillouin zone (BZ) at $k^{\pm}_{y}=\pm\cos^{-1}\frac{\beta-m}{\beta}$ for each value of $k_z.$ This system is in a non-generic LTSM phase with two straight FLs that traverse the entire BZ.  The nodes are locally stable in the BZ as long as the composite $\mathcal{T}\mathcal{I}$-symmetry is preserved. In fact,  in terms of the flux of the adiabatic connection, we note that the Dirac nodes carry a Berry flux of $\pm \pi,$  and hence are stabilized from the formation of an energy gap since doing so would smoothly spread the flux around the gapped degeneracy point, which is forbidden by $\mathcal{TI}$ symmetry\cite{bernevigbook,TSMresponse}. The FLs carry a helicity $\chi,$ which along with $\mbox{sgn}\,m_{{\cal{I}}},$ indicates the sign of the Berry flux. The response theory of this simple case can be directly determined from the results of Ref. \onlinecite{TSMresponse}.  Indeed, the response is given by
%\be
%S_{eff}[A_{\mu},b_{\nu}]=\frac{e}{(2\pi)^2}\int dt d^{3}x\,\epsilon^{\mu\nu\rho\sigma}A_{\mu}\del_{\nu}b_{\rho}\nu_{\sigma}
%\ee\noindent where $b_{\mu}$ is the intrinsic one-form determined by the (identical) nodal structure in any $(k_x,k_y)$-plane, and $\nu_{\sigma}=2\pi \delta_{\sigma z}$ is the reciprocal lattice vector normal to the stacked Dirac semimetal planes. The spatial components $b_{i}$ determine the polarization while the temporal component $b_{0}$ determines the magnetization via $eb_{\mu}=2\pi \mbox{sgn}\,m_{A}(M,\epsilon_{ij}P^{j})$. %The issue of turning on an inversion breaking mass to determine the sign of the polarization/magnetization is carried over to the 2D system as well. 
Eq. \ref{eq:action} with ${\cal{B}}=\mbox{sgn}\,m_{{\cal{I}}}(b_y\wedge {\bf{G}}_z)$ where $2b_y=k^{+}_{y}-k^{-}_{y}.$ 

%When both $\beta$ and $\gamma$ are very weak $\beta,\gamma\ll 1$, the system represents a secondary weak topological phase protected by translation and inversion symmetry. The secondary weak TIs have a two-index antisymmetric tensor topological invariant $\nu_{ij},$ where $i,j=1\ldots d$ of which only $\nu_{yz}=-\nu_{zy}$ are non-zero for the model $H_3.$
%To define polarization or magnetization for the sample as we will do in the following sections, we have to keep in mind that this mass term must be turned on for a finite size sample to fill the system appropriately and must be set to zero only after taking the system size to infinity just like we do in any case of spontaneous symmetry breaking. 
Now let us consider a more generic/isotropic case by increasing the tunneling strengths $\gamma, \beta$ large enough so that the insulating gap closes and a single closed FL inside the BZ forms. The gapless semimetallic phases of $H_3$ can be found using the constraints that $k_{x}=0$ or $\pi$ and $(1\mp 1+\beta+\gamma-m)=\beta \cos k_y+\gamma \cos k_z.$ For intuition we can consider the continuum limit and search for solutions near special points in the BZ. First let us consider an expansion around the origin where the constraint is $\beta k_{y}^2+\gamma k_{z}^2=2m,$ i.e., the equation for an ellipse.  Assuming that $\beta, \gamma>0$ to be explicit, this constraint only has a solution when $m>0.$ 
%We can also expand around $(k_x,k_{y},k_{z})=(0,\pi,\pi)$. This leads to the continuum-limit constraint  $\beta(\pi-\delta k_{y})^{2}+\gamma(\pi-\delta k_{z})^{2}=4+4\beta+4\gamma-2m$ where $\delta k_y, \delta k_z$ are the deviations from $\pi.$ This constraint is important to consider when $2+2\beta+2\gamma-m>0$ and small. 
Now to be concrete take $\beta=\gamma=2m=2$  so that there is a only a single FL circle located in the $k_x=0$ plane and none at $k_x=\pi.$ To illustrate the nature of the FL let us expand the Hamiltonian near it. It is convenient to switch to cylindrical coordinates: $(k_x,k_y,k_z)\to (k_x, \kappa, \theta)$ where $\theta$ winds around the FL and $\kappa$ represents the (signed) radial distance \emph{away} from the FL in the $k_yk_z$-plane. Using this definition, $(k_x=0,\kappa=0,\theta)$ lies on the FL and we find the Hamiltonian near the FL is
\begin{equation}
\label{eq:dirac_fs}
H_{FL}(k)\approx \delta k_x \sigma^y+(1/2(\delta k_x)^2+2\delta\kappa)\sigma^z\approx \delta k_x \sigma^x+m(\delta\kappa)\sigma^z
\end{equation}\noindent where the mass function $m(\delta\kappa)\equiv 2\delta\kappa$ and nothing depends on $\theta.$  Thus, near the Fermi surface we find a family of 1D Dirac Hamiltonians along the $x$-direction with masses depending on the radius in k-space away from the Fermi-surface $(\delta \kappa)$ in the $k_y k_z$-plane which can be positive or negative. This expansion shows that at each $(k_y,k_z)$ we have the Hamiltonian of a massive 1D Dirac model, and the sign of the mass  (and thus topological phase) changes as a function of $(k_y,k_z)$ as one passes through the FL. The 1D Bloch Hamiltonians $H_{k_y,k_z}(k_x)$ along $k_x$ (and parameterized by $(k_y,k_z)$) will be topologically trivial on one side of the FL and non-trivial on the other side. Which side is topological, and which is trivial depends not only on the nodal lines, but also on the secondary weak invariant of the filled bulk bands. This bulk weak invariant influences whether the surface states in the projected surface BZ are on the interior or exterior of the FL. %For the properties of the semimetal determined by the Fermi-surface alone, it does not matter which region is topological and which is trivial because they differ by the addition of completely filled bands carrying a secondary weak topological invariant.  

\begin{figure}
\label{fig:combined}
  \includegraphics[width=9cm]{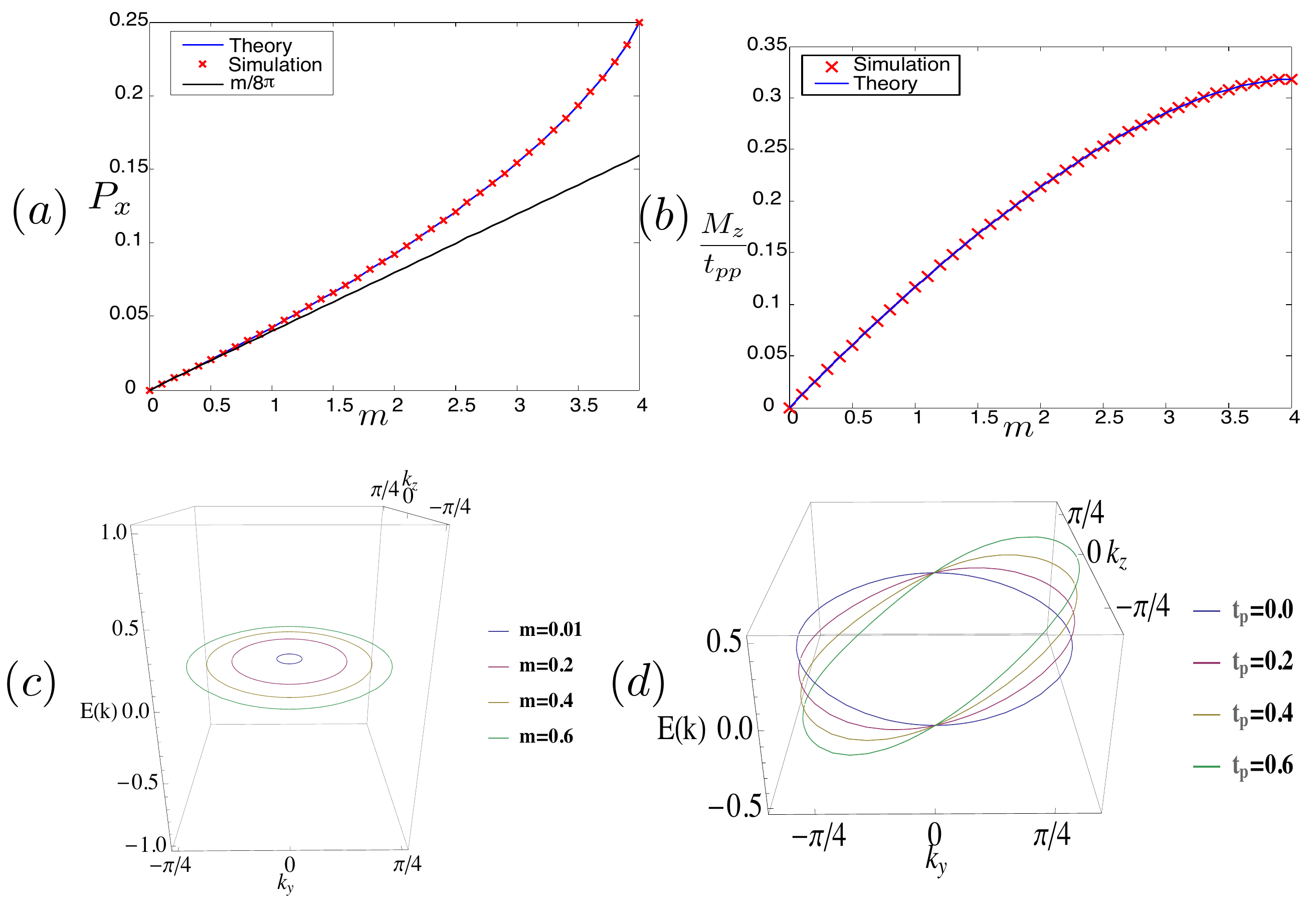}
   \caption{(a) The polarization for the model in Eq.\ref{eq:fermistring} is plotted vs the parameter $m$ in the model with $\beta=\gamma=2$. The polarization scales approximately linear with $m$ for small values of $m$, but deviates as $m$ is increased. (b) The magnetization for the model in Eq.\ref{eq:fermistring} with an extra term $t_{pp}\sin k_{y}\mathbb{I}$ is plotted vs various values of $m$ for $\beta=\gamma=2$. It is linear for small values of $m$, but ends up saturating when $m$ is increased. (c) The location of the line node is plotted in the $E-k_{y}-k_{z}$ space with $k_{x}=0$ for $t_{p}=0.0$  and various values of $m$ with $\beta=\gamma=2$. The polarization is proportional to the area enclosed by the FL. (d) The location of the line node is plotted in the $E-k_{y}-k_{z}$ space with $k_{x}=0$ for $m=1$  and various values of $t_{p}$ with $\beta=\gamma=2$. The magnetization is proportional to integral of the energy around the FL in momentum space.}
\end{figure}

As we have seen in the special case above ($\gamma=0, \beta\neq 0$), one quasi-topological EM response determined by the geometry of the FLs is the charge polarization. Let us consider this more generally.  
%To derive the precise results we begin by considering the charge polarization of a single FL. 
 The theory of electric polarization for insulators with gapped surfaces\cite{vanderbilt1993} was extended to Chern insulators with gapless edges in  Ref. \onlinecite{vanderbiltchern}, and 2D Dirac TSMs in Ref. \onlinecite{TSMresponse}.  In the latter two articles it was shown that insulators and TSMs with gapless boundary states can have a well-defined polarization,  though the the connection between the bulk periodic calculation of the polarization and the boundary charge requires care in handling the filling of the boundary modes. For a single FL, the boundary modes can be dealt with straightforwardly, but when there are multiple FLs the possibility of  $Z_2$ cancellations of overlapping surface states complicates the connection between the bulk value of the polarization (derived from Berry phase arguments below) and the surface charge\cite{TSMresponse}. This can be dealt with systematically (as was done for 2D Dirac semimetals in Ref. \onlinecite{TSMresponse}), and we resolve the case of two FLs in the Supporting Online Material.
%\footnote{The reliability of these predictions when the system is strongly disordered and not in the crystalline limit is unknown, though the same can be said for essentially all quasi-topological semimetal responses.}

To generally determine the polarization of a LTSM in some fixed direction $\hat{n}$ we consider the family of 1D Bloch Hamiltonians $H_{\vec{k}_{\perp}}(k_{\parallel})$ parameterized by $\vec{k}_{\perp},k_{\parallel}$, the components of the momentum perpendicular and parallel to $\hat{n}$. Generically the family $H_{\vec{k}_{\perp}}(k_{\parallel})$ is a set of 1D gapped Bloch Hamiltonians except when the point $(k_{\parallel},\vec{k}_{\perp})$ lies on one of the Dirac FLs (which only occupy a set of measure zero in the 3D BZ).  Note that while the FLs in our model are planar, our results below apply to non-planar cases as well.
 To calculate the charge polarization we first need to calculate the quantity\cite{vanderbilt1993}
\be
\Theta_{\parallel}(\vec{k}_{\perp})=\frac{e}{2\pi}\int dk_{\parallel}\mbox{Tr}\left[\,\mathcal{A}_{\parallel}(k_{\parallel},\vec{k}_{\perp})\right]
\ee\noindent where $\mathcal{A}_{\parallel}$ is the component of the Berry connection along $\hat{n}.$ Let us first consider the special case where we evaluate $\Theta (\vec{k}_{\perp})$ at $\vec{k}_{\perp}=\vec{\Lambda}_a$, where $\vec{\Lambda}_a$ is any inversion-invariant momentum in the $\vec{k}_\perp$ plane, i.e., $\vec{\Lambda}_a=-\vec{\Lambda}_a\mod \vec{G}.$ Then $\Theta(\vec{\Lambda_a})=0\mod e$ or  $\tfrac{e}{2}\mod e$ if $H_{\vec{\Lambda}_a}(k_\parallel)$ is gapped, since this 1D Bloch Hamiltonian has inversion symmetry. We then consider a deviation away from $\vec{k}_{\perp}=\Lambda_a$  which is still in the plane normal to $\hat{n},$ and such that the Hamiltonian $H_{\vec{\Lambda}_a+\vec{\delta k}_{\perp}}(k_\parallel)$ is gapped. However, this 1D Bloch Hamiltonian does not have to be inversion invariant, and thus it is not immediately obvious how to evaluate $\Theta_{\parallel}.$ However, we can use the following general argument to simplify the calculation. Let us evaluate the difference in the 1D polarizations
\begin{eqnarray}
\Delta \Theta_{\parallel}&=&\Theta_{\parallel}(\vec{\Lambda}_a+\vec{\delta k}_{\perp})-\Theta_{\parallel}(\vec{\Lambda}_a)\nonumber\\
&=&\frac{e}{2\pi}\int dk_{\parallel}\left\{\mbox{Tr}\left[\,\mathcal{A}_{\parallel}(k_{\parallel},\vec{\Lambda}_{a}+\vec{\delta k}_{\perp})\right]\right.\nonumber\\
&-&\left.\mbox{Tr}\left[\,\mathcal{A}_{\parallel}(k_{\parallel},\vec{\Lambda}_{a})\right]\right\}
=\frac{e}{2\pi}\int_S  \mbox{Tr}\left[F\right]
\end{eqnarray}\noindent where the last expression is a surface integral of the Berry curvature $2$-form $F$ over the region $S$ bounded by the two closed circles located at $\vec{\Lambda}_a$ and $\vec{\Lambda}_a+\vec{\delta k}_\perp$  and spanned by $k_{\parallel}$ through the cycle of the BZ in the $\hat{n}$ direction. Since our system has $\mathcal{T}\mathcal{I}$ symmetry, the only sources of Berry curvature are the $\pi$-flux lines carried by the Dirac FLs. Thus, generically $\Delta \Theta_{\parallel}= 0$ or $\tfrac{e}{2}$ depending on the parity of the number of Dirac line-nodes enclosed in the surface $S.$ In fact, this argument is completely general and does not rely on starting at an inversion-invariant momentum: $\Delta \Theta_{\parallel}=0$ or $\tfrac{e}{2}$ only relies on the existence of $\mathcal{T}\mathcal{I}$ symmetry. The ability to start at an inversion-invariant momentum informs us that the global constant needed to determine the full $\Theta_\parallel(\vec{k}_\perp)$ from the knowledge of only the $\Delta\Theta_\parallel(\vec{k}_{\perp})$ is either $0$ or $\tfrac{e}{2}$; data which is encoded in the secondary weak invariant $\nu_{ij}$ if $\Lambda_a$ is not the $\Gamma$-point. For a single FL we see that ${\mathcal{B}}_{ij},$ and hence, the overall charge polarization is simply proportional to the projected area of the FL in the $\hat{n}$ boundary BZ, i.e.,
\be
\label{eq:string_pol}
e\mathcal{B}_{ij}=4\pi^2\epsilon^{\hat{n}ij}P_{\hat{n}}=\int_{{\perp}BZ}d\vec{k}_{\perp}\Theta_{\parallel}(\vec{k}_\perp)=(-1)^{\nu_{ij}}\frac{e}{2}\Xi\,\Omega_{ij}
\ee
where $\Omega_{ij}$ is the area of the FL projected onto the boundary BZ, $\Xi=\chi(\mbox{sgn}\,m_{{\cal{I}}}),$  and $\chi=\pm1$ corresponds to the FL helicity, i.e., the clockwise/counterclockwise flow of the Berry flux along the FL with respect to the normal $\hat{n}$.  This bulk result holds up to the addition of a quantum of polarization\cite{vanderbilt1993}. Also, changing the secondary weak invariant $\nu_{ij},$  can change the polarization by a quantum, and/or a sign, since it switches the projected area to its complement in the surface BZ. For a single FL the effects are already taken into account in Eq. \ref{eq:string_pol}. For more than one FL, the bulk calculation will result in the projected areas of all the FLs modulo regions where an even number of FLs have overlapping projections.  As shown in the Supplementary Material, when FLs have overlapping projected areas the connection between this bulk result and the surface charge requires some knowledge of the filling of the boundary states.  

When an infinitesimal inversion-symmetry breaking mass $m_{{\cal{I}}}$ is added to the system, the surface states on one side will be filled while the other will be empty. Each filled surface state will contribute $\tfrac{e}{2}$ charge to the boundary which exactly matches the bound charge required from the polarization calculation. We confirm this result numerically in Fig.~\ref{fig:combined}a where $P_x$ of $H_3(k)$ is plotted vs. $m$ with the corresponding location of the FL shown in Fig.~\ref{fig:combined}c. We choose $\beta,\gamma=2$ so that there is a single FL in the $k_{x}=0$ plane and centered around the origin of the BZ. $P_x$ should be proportional to the area enclosed by the FL given by $\cos k_y+\cos k_z=2-m/2.$
For small values of $m$, the FL is approximately a circle of radius $\sqrt{m}$ and $P_{x}\approx\mbox{sgn}\,m_{{\cal{I}}}\frac{m}{8\pi}$. This approximation works well when $m$ is small, but underestimates $P_x$ as $m$ is increased. At $m=4$, the FL given by $\cos k_{y}+\cos k_{z}=0$ will enclose half the area of the BZ. We see that the polarization will have the symmetry $P_{x}(m)=\frac{e}{2}-P_{x}(8-m)$ simply because when $m>4$, the FL is centered around $(\pi,\pi)$ on the boundary BZ. Hence, we will restrict ourselves to $0\leq m\leq 4$ in Fig.~\ref{fig:combined}a.

One corollary of these general arguments is that, while it is not forbidden to have just a single closed FL in systems with $\mathcal{T}\mathcal{I}$ symmetry, it is forbidden to have only one (or an odd number) FLs which traverses a non-trivial cycle of the BZ and meets itself. We can see this because calculating any component of the polarization would indicate that the polarization must be opposite on either side of the FL, however this is not compatible with the periodicity of the BZ and thus must be forbidden. This is a 3D line-node generalization of the Fermion doubling theorem for Dirac nodes in 2D with $\mathcal{T}\mathcal{I}$ symmetry. 

%\begin{figure}
%\label{fig:contour}
%\includegraphics[width=9cm]{line_contourm.pdf}
%\caption{The location of the line node is plotted in the $E-k_{y}-k_{z}$ space with $k_{x}=0$ for $t_{p}=0.0$  and various values of $m$ with $\beta=\gamma=2$. The polarization is proportional to the area enclosed by the Fermi XXXX.}
%\end{figure}
%\begin{figure}
%\label{fig:polplot}
%\includegraphics[width=9cm]{fs_pol.pdf}
 % \caption{The polarization density in the $x$ direction is plotted for various values of $m$ when $\beta=\gamma=2$. The corresponding locations of the line node in $E-k_{y}-k_{z}$ space with $k_{x}=0$ is shown in Fig.~\ref{fig:contour}. A small inversion mass is turned on and periodic boundary conditions are assumed in the $y,z$ directions and the sample is a cube with 40 sites in every direction. The simulation values are compared with the approximate value for the polarization of $\frac{m}{8\pi}$ in blue and the exact calculation using the area in black. We note that the actual polarization is slightly larger as $m$ increases and matches it quite well at low $m$. }
%\end{figure}

Similar to the 2D Dirac TSMs, which have a non-vanishing orbital magnetization when there is an energy difference between the Dirac nodes, LTSMs can also have a magnetization that depends on how the band touching lines are embedded in energy/momentum space. To produce this effect in our model we need to change the energy along the nodal submanifold, and we can do this, e.g.,  by adding an extra kinetic energy term  $\epsilon(\vec{k})\mathbb{I}$ to $H_3(k).$  Following Refs. \onlinecite{niu2007,vanderbiltmag}, we have
\bea
M^{a}&=&\frac{e\epsilon^{abc}}{2\hbar}\int \frac{d^3k}{(2\pi)^3}{\rm{Im}}\langle\partial_b u_{-}\vert( H_3(k)+\\\nonumber
&&\epsilon(k)+E_{-}(k))\vert\partial_c u_{-}\rangle
\eea\noindent where $E_{-}(k), \vert u_{-}\rangle$ are the energy and Bloch functions of the lower occupied band, and the derivatives are with respect to momentum. This is evaluated in detail in the Supplementary Material. The main property that simplifies this computation is that the Berry curvature is composed of just $\pi$-flux $\delta$-functions due to the $\mathcal{TI}$ symmetry. We find\be
\label{eq:string_mag}
\frac{e\mathcal{B}^{0a}}{4\pi^2}=M^a=\frac{e\Xi}{4\pi\hbar}\int_{\del R}\epsilon(\vec{k})dk^a
\ee\noindent where the integration is over the nodal line.  Similar to the magnetization in the 2D Dirac semimetal, the resulting $\vec{M}$ does not depend on the weak invariant $\nu_{ij}$. In a generic model, Eq. \ref{eq:string_mag} will include a sum over integrals for all distinct FLs. This is a 3D generalization of the results of Refs. \onlinecite{niuvalley,TSMresponse} that relate the magnetization of the 2D Dirac semimetal to the energy differences between the band-touching points. 

We confirm this result numerically by adding an extra term $t_{pp}\sin k_{y}\mathbb{I}$ to $H_3(k)$ and plotting the magnetization vs. $m$ in Fig.~\ref{fig:combined}b. Again we have fixed $\beta,\gamma=2$ so that there is only one FL, which has $\chi=+1.$  The magnetization for this case can be evaluated analytically from Eq.~\ref{eq:string_mag} since the energy only depends on $k_y.$ The limits to which $k_{z}$ extends for the FL can be calculated using the equation for the nodal line ($\cos k_{y}+\cos k_z=2-m/2$).  Hence,  on the nodal line, $k_{y}$ is a function of $k_{z}$. The maximum value of $\cos k_y=1$ and this means that the maximum/minimum $k_{z}$ is given by $\pm k_{z0}=\pm \cos^{-1}(1-m/2)$. This is valid only when $m<4$, while for $m>4$, the FL is centered around $(\pi,\pi)$ instead of the origin. The magnetization will have the symmetry $M_{z}(m)=M_{z}(8-m)$ which is why we restrict ourselves to $0\leq m\leq 4$ in Fig.~\ref{fig:combined}b. The magnetization is given by 
\bea
\frac{M^{z}(m)}{t_{pp}}&=&\mbox{sgn}\,m_{{\cal{I}}}\frac{e}{4\pi \hbar}\int_{-k_0}^{k_0} \sin k_{y}\,dk_{z}.
\eea\noindent The magnetization is a function of $m$ and does not have a simple closed form expression, but has a linear profile in the regime when $m$ is small. For surfaces with low-energy modes we can give a microscopic argument for the existence of the magnetization. The surface states of $H_3(k)$ are initially flat-bands that do not disperse and $\epsilon (k)$ will impart a dispersion as a function of $(k_{y},k_{z})$. In general, this will create a bound surface current in the $y-z$ plane which is the consequence of a non-vanishing bulk magnetization density. There will be similar currents on surfaces without low-energy modes, but there is not as simple of an interpretation\cite{TSMresponse}.
%\begin{figure}
%\includegraphics[width=9cm]{line_contour.pdf}
%\caption{The location of the line node is plotted in the $E-k_{y}-k_{z}$ space with $k_{x}=0$ for $m=1$  and various values of $t_{p}$ with $\beta=\gamma=2$. The magnetization is proportional to integral of the energy around the Fermi XXXX in momentum space.}
%\end{figure}

%\begin{figure}
%\centering
%\label{fig:magplot}
 %  \includegraphics[width=9cm]{fs_mag.pdf}
%   \caption{The magnetization in the $y$ direction relative to the coefficient $t_{pp}$ is plotted for various values of $m$ when $\beta=\gamma=2$. A small inversion mass is turned on and periodic boundary conditions are assumed in the $y,z$ directions and the sample is a cube of size $40\times40\times40$. The simulation values are compared with what is predicted by Eq.~\ref{eq:string_mag}. }
%\end{figure}

\begin{figure}
\label{fig:split}
  \includegraphics[width=8.8cm]{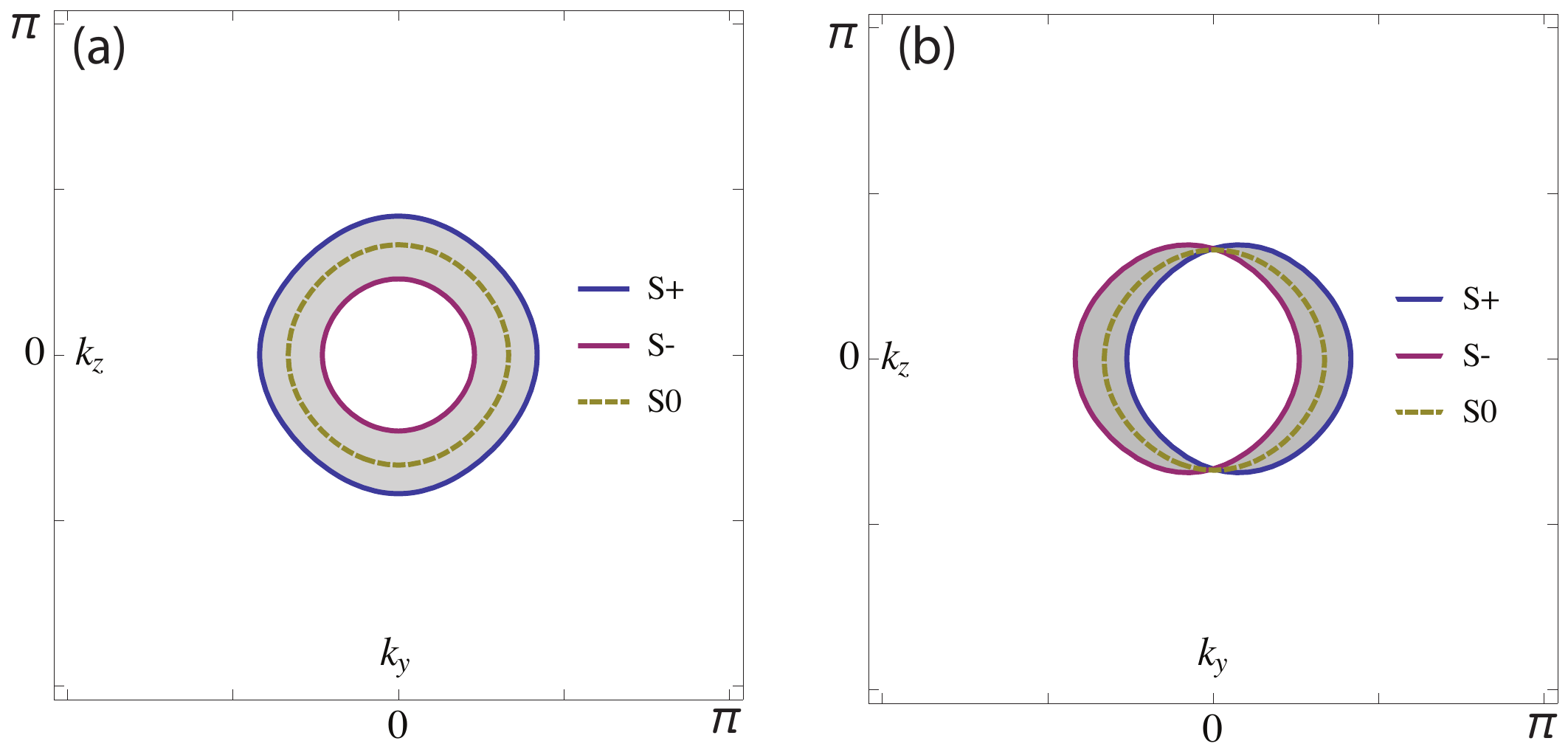}
   \caption{Dotted yellow lines represent initial four-fold degenerate Fermi-line ($S0$). Purple and blue solid lines represent spin-split Fermi-surfaces $(S+,S-)$ with (a) a majority and minority spin Fermi-line induced by certain ${\cal{T}}$-breaking terms (b) spin-split Fermi-lines with equal sizes for each spin reminiscent of a Rashba-type splitting from spin-orbit terms induced by strain/inversion breaking. For both panels the gray shaded region represents the magnitude of the polarization in the $x$-direction from the projected areas of the Fermi lines after $Z_2$ overlap cancellation. }
\end{figure}

We have now completed our goal of showing that the FL EM response is given by Eq. 1. To conclude,  we comment on the applicability of our results to real materials. The magnetic heterostructure proposed in Refs.~\onlinecite{burkov2011,phillips2014} breaks ${\cal{T}}$ explicitly, hence the spins are not degenerate and the line nodes occur with just two overlapping bands.  Thus, this model %the line nodes are protected under mirror symmetries about the $x,y,z$ directions and they  
corresponds precisely to an effectively spinless case that has been described throughout this paper and our results can be directly applied. We expect, and have confirmed numerically, that this system will have a charge polarization. %when projected onto one of the sub bands as done in the original model as shown in Sec. 4B2 of Ref.~\onlinecite{burkov2011}. 
 In the case of spin degenerate models, which are found, for example, in the carbon allotrope materials in Refs.~\onlinecite{uchoa2014, weng2014}, a further reflection symmetry is required to stabilize the LTSM arising from four overlapping bands as shown in Ref. \onlinecite{fang2015}. For doubly-degenerate bands the charge polarization, being a $\mathbb{Z}_{2}$ quantity, is trivial. However, we can break spin degeneracy by including certain ${\cal{T}}$-breaking terms,  or inducing additional spin orbit terms via strain, with the requirement that the FLs are not completely destabilized to a gapped, or point-node, phase. If we take two copies of our model, one for each spin, then two illustrations of initially spin-degenerate FLs  (in the $k_x=0$ plane) split by two types of spin-dependent terms are shown in Fig. \ref{fig:split}. In these cases, the polarization $P_x$ can be nontrivial and is not completely $Z_2$ canceled. In fact, in both cases, the shaded areas correspond to the magnitude of the polarization, assuming a vanishing secondary weak invariant. The magnetization, on the other hand, is not a $Z_2$ quantity and can be non-vanishing even for four-fold degenerate FLs. Hence, we expect that these systems would exhibit charge polarization when the FLs are spin-split via strain or other spin-dependent perturbations. 

\acknowledgements{We would like to thank P.Y. Chang, V.K. Chua, V. Dwivedi, and A. Tiwari for discussions. We acknowledge support from ONR YIP Award N00014-15-1-2383.}
%\bibliographystyle{apsrev4-1}
%\bibliography{InteractingWeakBib.bib}

\begin{thebibliography}{67}%
\makeatletter
\providecommand \@ifxundefined [1]{%
 \@ifx{#1\undefined}
}%
\providecommand \@ifnum [1]{%
 \ifnum #1\expandafter \@firstoftwo
 \else \expandafter \@secondoftwo
 \fi
}%
\providecommand \@ifx [1]{%
 \ifx #1\expandafter \@firstoftwo
 \else \expandafter \@secondoftwo
 \fi
}%
\providecommand \natexlab [1]{#1}%
\providecommand \enquote  [1]{``#1''}%
\providecommand \bibnamefont  [1]{#1}%
\providecommand \bibfnamefont [1]{#1}%
\providecommand \citenamefont [1]{#1}%
\providecommand \href@noop [0]{\@secondoftwo}%
\providecommand \href [0]{\begingroup \@sanitize@url \@href}%
\providecommand \@href[1]{\@@startlink{#1}\@@href}%
\providecommand \@@href[1]{\endgroup#1\@@endlink}%
\providecommand \@sanitize@url [0]{\catcode `\\12\catcode `\$12\catcode
  `\&12\catcode `\#12\catcode `\^12\catcode `\_12\catcode `\%12\relax}%
\providecommand \@@startlink[1]{}%
\providecommand \@@endlink[0]{}%
\providecommand \url  [0]{\begingroup\@sanitize@url \@url }%
\providecommand \@url [1]{\endgroup\@href {#1}{\urlprefix }}%
\providecommand \urlprefix  [0]{URL }%
\providecommand \Eprint [0]{\href }%
\providecommand \doibase [0]{http://dx.doi.org/}%
\providecommand \selectlanguage [0]{\@gobble}%
\providecommand \bibinfo  [0]{\@secondoftwo}%
\providecommand \bibfield  [0]{\@secondoftwo}%
\providecommand \translation [1]{[#1]}%
\providecommand \BibitemOpen [0]{}%
\providecommand \bibitemStop [0]{}%
\providecommand \bibitemNoStop [0]{.\EOS\space}%
\providecommand \EOS [0]{\spacefactor3000\relax}%
\providecommand \BibitemShut  [1]{\csname bibitem#1\endcsname}%
\let\auto@bib@innerbib\@empty
%</preamble>
\bibitem [{\citenamefont {Hasan}\ and\ \citenamefont {Kane}(2010)}]{HasanKane}%
  \BibitemOpen
  \bibfield  {author} {\bibinfo {author} {\bibfnamefont {M.~Z.}\ \bibnamefont
  {Hasan}}\ and\ \bibinfo {author} {\bibfnamefont {C.~L.}\ \bibnamefont
  {Kane}},\ }\href@noop {} {\bibfield  {journal} {\bibinfo  {journal} {Rev.
  Mod. Phys.}\ }\textbf {\bibinfo {volume} {82}},\ \bibinfo {pages} {3045}
  (\bibinfo {year} {2010})}\BibitemShut {NoStop}%
\bibitem [{\citenamefont {Bernevig}(2013)}]{bernevigbook}%
  \BibitemOpen
  \bibfield  {author} {\bibinfo {author} {\bibfnamefont {B.~A.}\ \bibnamefont
  {Bernevig}},\ }\href@noop {} {\emph {\bibinfo {title} {Topological Insulators
  and Topological Superconductors}}}\ (\bibinfo  {publisher} {Princeton
  University Press},\ \bibinfo {year} {2013})\BibitemShut {NoStop}%
\bibitem [{\citenamefont {Haldane}(1988)}]{haldane1988}%
  \BibitemOpen
  \bibfield  {author} {\bibinfo {author} {\bibfnamefont {F.}~\bibnamefont
  {Haldane}},\ }\href@noop {} {\bibfield  {journal} {\bibinfo  {journal} {Phys.
  Rev. Lett.}\ }\textbf {\bibinfo {volume} {61}},\ \bibinfo {pages} {2015}
  (\bibinfo {year} {1988})}\BibitemShut {NoStop}%
\bibitem [{\citenamefont {Qi}\ \emph {et~al.}(2008{\natexlab{a}})\citenamefont
  {Qi}, \citenamefont {Hughes},\ and\ \citenamefont {Zhang}}]{qi2008}%
  \BibitemOpen
  \bibfield  {author} {\bibinfo {author} {\bibfnamefont {X.-L.}\ \bibnamefont
  {Qi}}, \bibinfo {author} {\bibfnamefont {T.~L.}\ \bibnamefont {Hughes}}, \
  and\ \bibinfo {author} {\bibfnamefont {S.-C.}\ \bibnamefont {Zhang}},\
  }\href@noop {} {\bibfield  {journal} {\bibinfo  {journal} {Phys. Rev. B}\
  }\textbf {\bibinfo {volume} {78}},\ \bibinfo {pages} {195424} (\bibinfo
  {year} {2008}{\natexlab{a}})}\BibitemShut {NoStop}%
\bibitem [{\citenamefont {Schnyder}\ \emph {et~al.}(2008)\citenamefont
  {Schnyder}, \citenamefont {Ryu}, \citenamefont {Furusaki},\ and\
  \citenamefont {Ludwig}}]{schnyder2008}%
  \BibitemOpen
  \bibfield  {author} {\bibinfo {author} {\bibfnamefont {A.~P.}\ \bibnamefont
  {Schnyder}}, \bibinfo {author} {\bibfnamefont {S.}~\bibnamefont {Ryu}},
  \bibinfo {author} {\bibfnamefont {A.}~\bibnamefont {Furusaki}}, \ and\
  \bibinfo {author} {\bibfnamefont {A.~W.}\ \bibnamefont {Ludwig}},\
  }\href@noop {} {\bibfield  {journal} {\bibinfo  {journal} {Phys. Rev. B}\
  }\textbf {\bibinfo {volume} {78}},\ \bibinfo {pages} {195125} (\bibinfo
  {year} {2008})}\BibitemShut {NoStop}%
\bibitem [{\citenamefont {Kitaev}(2009)}]{kitaev2009}%
  \BibitemOpen
  \bibfield  {author} {\bibinfo {author} {\bibfnamefont {A.}~\bibnamefont
  {Kitaev}},\ }\href@noop {} {\bibfield  {journal} {\bibinfo  {journal} {arXiv
  preprint arXiv:0901.2686}\ } (\bibinfo {year} {2009})}\BibitemShut {NoStop}%
\bibitem [{\citenamefont {Fu}\ and\ \citenamefont
  {Kane}(2007)}]{FuKaneWeak2007}%
  \BibitemOpen
  \bibfield  {author} {\bibinfo {author} {\bibfnamefont {L.}~\bibnamefont
  {Fu}}\ and\ \bibinfo {author} {\bibfnamefont {C.~L.}\ \bibnamefont {Kane}},\
  }\href@noop {} {\bibfield  {journal} {\bibinfo  {journal} {Phys. Rev. B}\
  }\textbf {\bibinfo {volume} {76}},\ \bibinfo {pages} {045302} (\bibinfo
  {year} {2007})}\BibitemShut {NoStop}%
\bibitem [{\citenamefont {Teo}\ \emph {et~al.}(2008)\citenamefont {Teo},
  \citenamefont {Fu},\ and\ \citenamefont {Kane}}]{teo2008}%
  \BibitemOpen
  \bibfield  {author} {\bibinfo {author} {\bibfnamefont {J.~C.~Y.}\
  \bibnamefont {Teo}}, \bibinfo {author} {\bibfnamefont {L.}~\bibnamefont
  {Fu}}, \ and\ \bibinfo {author} {\bibfnamefont {C.~L.}\ \bibnamefont
  {Kane}},\ }\href@noop {} {\bibfield  {journal} {\bibinfo  {journal} {Phys.
  Rev. B}\ }\textbf {\bibinfo {volume} {78}},\ \bibinfo {pages} {045426}
  (\bibinfo {year} {2008})}\BibitemShut {NoStop}%
\bibitem [{\citenamefont {Fu}(2011)}]{fu2011topological}%
  \BibitemOpen
  \bibfield  {author} {\bibinfo {author} {\bibfnamefont {L.}~\bibnamefont
  {Fu}},\ }\href@noop {} {\bibfield  {journal} {\bibinfo  {journal} {Phys. Rev.
  Lett.}\ }\textbf {\bibinfo {volume} {106}},\ \bibinfo {pages} {106802}
  (\bibinfo {year} {2011})}\BibitemShut {NoStop}%
\bibitem [{\citenamefont {Hughes}\ \emph
  {et~al.}(2011{\natexlab{a}})\citenamefont {Hughes}, \citenamefont {Prodan},\
  and\ \citenamefont {Bernevig}}]{hughes2011inversion}%
  \BibitemOpen
  \bibfield  {author} {\bibinfo {author} {\bibfnamefont {T.~L.}\ \bibnamefont
  {Hughes}}, \bibinfo {author} {\bibfnamefont {E.}~\bibnamefont {Prodan}}, \
  and\ \bibinfo {author} {\bibfnamefont {B.~A.}\ \bibnamefont {Bernevig}},\
  }\href@noop {} {\bibfield  {journal} {\bibinfo  {journal} {Phys. Rev. B}\
  }\textbf {\bibinfo {volume} {83}},\ \bibinfo {pages} {245132} (\bibinfo
  {year} {2011}{\natexlab{a}})}\BibitemShut {NoStop}%
\bibitem [{\citenamefont {Turner}\ \emph {et~al.}(2012)\citenamefont {Turner},
  \citenamefont {Zhang}, \citenamefont {Mong},\ and\ \citenamefont
  {Vishwanath}}]{turner2012}%
  \BibitemOpen
  \bibfield  {author} {\bibinfo {author} {\bibfnamefont {A.~M.}\ \bibnamefont
  {Turner}}, \bibinfo {author} {\bibfnamefont {Y.}~\bibnamefont {Zhang}},
  \bibinfo {author} {\bibfnamefont {R.~S.}\ \bibnamefont {Mong}}, \ and\
  \bibinfo {author} {\bibfnamefont {A.}~\bibnamefont {Vishwanath}},\
  }\href@noop {} {\bibfield  {journal} {\bibinfo  {journal} {Physical Review
  B}\ }\textbf {\bibinfo {volume} {85}},\ \bibinfo {pages} {165120} (\bibinfo
  {year} {2012})}\BibitemShut {NoStop}%
\bibitem [{\citenamefont {Fang}\ \emph {et~al.}(2012)\citenamefont {Fang},
  \citenamefont {Gilbert},\ and\ \citenamefont {Bernevig}}]{fang2012bulk}%
  \BibitemOpen
  \bibfield  {author} {\bibinfo {author} {\bibfnamefont {C.}~\bibnamefont
  {Fang}}, \bibinfo {author} {\bibfnamefont {M.~J.}\ \bibnamefont {Gilbert}}, \
  and\ \bibinfo {author} {\bibfnamefont {B.~A.}\ \bibnamefont {Bernevig}},\
  }\href@noop {} {\bibfield  {journal} {\bibinfo  {journal} {Phys. Rev. B}\
  }\textbf {\bibinfo {volume} {86}},\ \bibinfo {pages} {115112} (\bibinfo
  {year} {2012})}\BibitemShut {NoStop}%
\bibitem [{\citenamefont {Teo}\ and\ \citenamefont
  {Hughes}(2013)}]{teohughes1}%
  \BibitemOpen
  \bibfield  {author} {\bibinfo {author} {\bibfnamefont {J.~C.}\ \bibnamefont
  {Teo}}\ and\ \bibinfo {author} {\bibfnamefont {T.~L.}\ \bibnamefont
  {Hughes}},\ }\href@noop {} {\bibfield  {journal} {\bibinfo  {journal} {Phys.
  Rev. Lett.}\ }\textbf {\bibinfo {volume} {111}},\ \bibinfo {pages} {047006}
  (\bibinfo {year} {2013})}\BibitemShut {NoStop}%
\bibitem [{\citenamefont {Slager}\ \emph {et~al.}(2012)\citenamefont {Slager},
  \citenamefont {Mesaros}, \citenamefont {Juri{\v{c}}i{\'c}},\ and\
  \citenamefont {Zaanen}}]{slager2012}%
  \BibitemOpen
  \bibfield  {author} {\bibinfo {author} {\bibfnamefont {R.-J.}\ \bibnamefont
  {Slager}}, \bibinfo {author} {\bibfnamefont {A.}~\bibnamefont {Mesaros}},
  \bibinfo {author} {\bibfnamefont {V.}~\bibnamefont {Juri{\v{c}}i{\'c}}}, \
  and\ \bibinfo {author} {\bibfnamefont {J.}~\bibnamefont {Zaanen}},\
  }\href@noop {} {\bibfield  {journal} {\bibinfo  {journal} {Nat. Phys.}\
  }\textbf {\bibinfo {volume} {9}},\ \bibinfo {pages} {98} (\bibinfo {year}
  {2012})}\BibitemShut {NoStop}%
\bibitem [{\citenamefont {Benalcazar}\ \emph {et~al.}(2013)\citenamefont
  {Benalcazar}, \citenamefont {Teo},\ and\ \citenamefont
  {Hughes}}]{Benalcazar2013}%
  \BibitemOpen
  \bibfield  {author} {\bibinfo {author} {\bibfnamefont {W.~A.}\ \bibnamefont
  {Benalcazar}}, \bibinfo {author} {\bibfnamefont {J.~C.}\ \bibnamefont {Teo}},
  \ and\ \bibinfo {author} {\bibfnamefont {T.~L.}\ \bibnamefont {Hughes}},\
  }\href@noop {} {\bibfield  {journal} {\bibinfo  {journal} {arXiv preprint
  arXiv:1311.0496}\ } (\bibinfo {year} {2013})}\BibitemShut {NoStop}%
\bibitem [{\citenamefont {Morimoto}\ and\ \citenamefont
  {Furusaki}(2013)}]{Morimoto2013}%
  \BibitemOpen
  \bibfield  {author} {\bibinfo {author} {\bibfnamefont {T.}~\bibnamefont
  {Morimoto}}\ and\ \bibinfo {author} {\bibfnamefont {A.}~\bibnamefont
  {Furusaki}},\ }\href@noop {} {\bibfield  {journal} {\bibinfo  {journal}
  {Phys. Rev. B}\ }\textbf {\bibinfo {volume} {88}},\ \bibinfo {pages} {125129}
  (\bibinfo {year} {2013})}\BibitemShut {NoStop}%
\bibitem [{\citenamefont {Chiu}\ \emph {et~al.}(2013)\citenamefont {Chiu},
  \citenamefont {Yao},\ and\ \citenamefont {Ryu}}]{RyuReflection2013}%
  \BibitemOpen
  \bibfield  {author} {\bibinfo {author} {\bibfnamefont {C.-K.}\ \bibnamefont
  {Chiu}}, \bibinfo {author} {\bibfnamefont {H.}~\bibnamefont {Yao}}, \ and\
  \bibinfo {author} {\bibfnamefont {S.}~\bibnamefont {Ryu}},\ }\href@noop {}
  {\bibfield  {journal} {\bibinfo  {journal} {Phys. Rev. B}\ }\textbf {\bibinfo
  {volume} {88}},\ \bibinfo {pages} {075142} (\bibinfo {year}
  {2013})}\BibitemShut {NoStop}%
\bibitem [{\citenamefont {Fang}\ \emph {et~al.}(2013)\citenamefont {Fang},
  \citenamefont {Gilbert},\ and\ \citenamefont
  {Bernevig}}]{fang2013entanglement}%
  \BibitemOpen
  \bibfield  {author} {\bibinfo {author} {\bibfnamefont {C.}~\bibnamefont
  {Fang}}, \bibinfo {author} {\bibfnamefont {M.~J.}\ \bibnamefont {Gilbert}}, \
  and\ \bibinfo {author} {\bibfnamefont {B.~A.}\ \bibnamefont {Bernevig}},\
  }\href@noop {} {\bibfield  {journal} {\bibinfo  {journal} {Phys. Rev. B}\
  }\textbf {\bibinfo {volume} {87}},\ \bibinfo {pages} {035119} (\bibinfo
  {year} {2013})}\BibitemShut {NoStop}%
\bibitem [{\citenamefont {Hughes}\ \emph
  {et~al.}(2013{\natexlab{a}})\citenamefont {Hughes}, \citenamefont {Yao},\
  and\ \citenamefont {Qi}}]{HughesYaoQi13}%
  \BibitemOpen
  \bibfield  {author} {\bibinfo {author} {\bibfnamefont {T.~L.}\ \bibnamefont
  {Hughes}}, \bibinfo {author} {\bibfnamefont {H.}~\bibnamefont {Yao}}, \ and\
  \bibinfo {author} {\bibfnamefont {X.-L.}\ \bibnamefont {Qi}},\ }\href@noop {}
  {\bibfield  {journal} {\bibinfo  {journal} {arXiv:1303.1539}\ } (\bibinfo
  {year} {2013}{\natexlab{a}})}\BibitemShut {NoStop}%
\bibitem [{\citenamefont {Ueno}\ \emph {et~al.}(2013)\citenamefont {Ueno},
  \citenamefont {Yamakage}, \citenamefont {Tanaka},\ and\ \citenamefont
  {Sato}}]{Sato2013}%
  \BibitemOpen
  \bibfield  {author} {\bibinfo {author} {\bibfnamefont {Y.}~\bibnamefont
  {Ueno}}, \bibinfo {author} {\bibfnamefont {A.}~\bibnamefont {Yamakage}},
  \bibinfo {author} {\bibfnamefont {Y.}~\bibnamefont {Tanaka}}, \ and\ \bibinfo
  {author} {\bibfnamefont {M.}~\bibnamefont {Sato}},\ }\href@noop {} {\bibfield
   {journal} {\bibinfo  {journal} {Phys. Rev. Lett.}\ }\textbf {\bibinfo
  {volume} {111}},\ \bibinfo {pages} {087002} (\bibinfo {year}
  {2013})}\BibitemShut {NoStop}%
\bibitem [{\citenamefont {Zhang}\ \emph {et~al.}(2013)\citenamefont {Zhang},
  \citenamefont {Kane},\ and\ \citenamefont {Mele}}]{Kane2013Mirror}%
  \BibitemOpen
  \bibfield  {author} {\bibinfo {author} {\bibfnamefont {F.}~\bibnamefont
  {Zhang}}, \bibinfo {author} {\bibfnamefont {C.}~\bibnamefont {Kane}}, \ and\
  \bibinfo {author} {\bibfnamefont {E.}~\bibnamefont {Mele}},\ }\href@noop {}
  {\bibfield  {journal} {\bibinfo  {journal} {Phys. Rev. Lett.}\ }\textbf
  {\bibinfo {volume} {111}},\ \bibinfo {pages} {056403} (\bibinfo {year}
  {2013})}\BibitemShut {NoStop}%
\bibitem [{\citenamefont {Jadaun}\ \emph {et~al.}(2013)\citenamefont {Jadaun},
  \citenamefont {Xiao}, \citenamefont {Niu},\ and\ \citenamefont
  {Banerjee}}]{jadaun2013}%
  \BibitemOpen
  \bibfield  {author} {\bibinfo {author} {\bibfnamefont {P.}~\bibnamefont
  {Jadaun}}, \bibinfo {author} {\bibfnamefont {D.}~\bibnamefont {Xiao}},
  \bibinfo {author} {\bibfnamefont {Q.}~\bibnamefont {Niu}}, \ and\ \bibinfo
  {author} {\bibfnamefont {S.~K.}\ \bibnamefont {Banerjee}},\ }\href@noop {}
  {\bibfield  {journal} {\bibinfo  {journal} {Phys. Rev. B}\ }\textbf {\bibinfo
  {volume} {88}},\ \bibinfo {pages} {085110} (\bibinfo {year}
  {2013})}\BibitemShut {NoStop}%
\bibitem [{\citenamefont {Hsieh}\ \emph {et~al.}(2008)\citenamefont {Hsieh},
  \citenamefont {Qian}, \citenamefont {Wray}, \citenamefont {Xia},
  \citenamefont {Hor}, \citenamefont {Cava},\ and\ \citenamefont
  {Hasan}}]{hsieh08}%
  \BibitemOpen
  \bibfield  {author} {\bibinfo {author} {\bibfnamefont {D.}~\bibnamefont
  {Hsieh}}, \bibinfo {author} {\bibfnamefont {D.}~\bibnamefont {Qian}},
  \bibinfo {author} {\bibfnamefont {L.}~\bibnamefont {Wray}}, \bibinfo {author}
  {\bibfnamefont {Y.}~\bibnamefont {Xia}}, \bibinfo {author} {\bibfnamefont
  {Y.~S.}\ \bibnamefont {Hor}}, \bibinfo {author} {\bibfnamefont {R.~J.}\
  \bibnamefont {Cava}}, \ and\ \bibinfo {author} {\bibfnamefont {M.~Z.}\
  \bibnamefont {Hasan}},\ }\href {\doibase 10.1038/nature06843} {\bibfield
  {journal} {\bibinfo  {journal} {Nature}\ }\textbf {\bibinfo {volume} {452}},\
  \bibinfo {pages} {970} (\bibinfo {year} {2008})}\BibitemShut {NoStop}%
\bibitem [{\citenamefont {Moore}(2009)}]{moore2009topological}%
  \BibitemOpen
  \bibfield  {author} {\bibinfo {author} {\bibfnamefont {J.}~\bibnamefont
  {Moore}},\ }\href@noop {} {\bibfield  {journal} {\bibinfo  {journal} {Nat.
  Phys.}\ }\textbf {\bibinfo {volume} {5}},\ \bibinfo {pages} {378} (\bibinfo
  {year} {2009})}\BibitemShut {NoStop}%
\bibitem [{\citenamefont {Xia}\ \emph {et~al.}(2009)\citenamefont {Xia},
  \citenamefont {Qian}, \citenamefont {Hsieh}, \citenamefont {Wray},
  \citenamefont {Pal}, \citenamefont {Lin}, \citenamefont {Bansil},
  \citenamefont {Grauer}, \citenamefont {Hor}, \citenamefont {Cava} \emph
  {et~al.}}]{xia2009observation}%
  \BibitemOpen
  \bibfield  {author} {\bibinfo {author} {\bibfnamefont {Y.}~\bibnamefont
  {Xia}}, \bibinfo {author} {\bibfnamefont {D.}~\bibnamefont {Qian}}, \bibinfo
  {author} {\bibfnamefont {D.}~\bibnamefont {Hsieh}}, \bibinfo {author}
  {\bibfnamefont {L.}~\bibnamefont {Wray}}, \bibinfo {author} {\bibfnamefont
  {A.}~\bibnamefont {Pal}}, \bibinfo {author} {\bibfnamefont {H.}~\bibnamefont
  {Lin}}, \bibinfo {author} {\bibfnamefont {A.}~\bibnamefont {Bansil}},
  \bibinfo {author} {\bibfnamefont {D.}~\bibnamefont {Grauer}}, \bibinfo
  {author} {\bibfnamefont {Y.}~\bibnamefont {Hor}}, \bibinfo {author}
  {\bibfnamefont {R.}~\bibnamefont {Cava}},  \emph {et~al.},\ }\href@noop {}
  {\bibfield  {journal} {\bibinfo  {journal} {Nat. Phys.}\ }\textbf {\bibinfo
  {volume} {5}},\ \bibinfo {pages} {398} (\bibinfo {year} {2009})}\BibitemShut
  {NoStop}%
\bibitem [{\citenamefont {Zhang}\ \emph {et~al.}(2009)\citenamefont {Zhang},
  \citenamefont {Cheng}, \citenamefont {Chen}, \citenamefont {Jia},
  \citenamefont {Ma}, \citenamefont {He}, \citenamefont {Wang}, \citenamefont
  {Zhang}, \citenamefont {Dai}, \citenamefont {Fang} \emph
  {et~al.}}]{zhang2009experimental}%
  \BibitemOpen
  \bibfield  {author} {\bibinfo {author} {\bibfnamefont {T.}~\bibnamefont
  {Zhang}}, \bibinfo {author} {\bibfnamefont {P.}~\bibnamefont {Cheng}},
  \bibinfo {author} {\bibfnamefont {X.}~\bibnamefont {Chen}}, \bibinfo {author}
  {\bibfnamefont {J.-F.}\ \bibnamefont {Jia}}, \bibinfo {author} {\bibfnamefont
  {X.}~\bibnamefont {Ma}}, \bibinfo {author} {\bibfnamefont {K.}~\bibnamefont
  {He}}, \bibinfo {author} {\bibfnamefont {L.}~\bibnamefont {Wang}}, \bibinfo
  {author} {\bibfnamefont {H.}~\bibnamefont {Zhang}}, \bibinfo {author}
  {\bibfnamefont {X.}~\bibnamefont {Dai}}, \bibinfo {author} {\bibfnamefont
  {Z.}~\bibnamefont {Fang}},  \emph {et~al.},\ }\href@noop {} {\bibfield
  {journal} {\bibinfo  {journal} {Phys. Rev. Lett.}\ }\textbf {\bibinfo
  {volume} {103}},\ \bibinfo {pages} {266803} (\bibinfo {year}
  {2009})}\BibitemShut {NoStop}%
\bibitem [{\citenamefont {Kane}\ and\ \citenamefont {Mele}(2005)}]{Kane2005A}%
  \BibitemOpen
  \bibfield  {author} {\bibinfo {author} {\bibfnamefont {C.~L.}\ \bibnamefont
  {Kane}}\ and\ \bibinfo {author} {\bibfnamefont {E.~J.}\ \bibnamefont
  {Mele}},\ }\href@noop {} {\bibfield  {journal} {\bibinfo  {journal} {Phys.
  Rev. Lett.}\ }\textbf {\bibinfo {volume} {95}},\ \bibinfo {pages} {226801}
  (\bibinfo {year} {2005})}\BibitemShut {NoStop}%
\bibitem [{\citenamefont {Bernevig}\ \emph {et~al.}(2006)\citenamefont
  {Bernevig}, \citenamefont {Hughes},\ and\ \citenamefont
  {Zhang}}]{bernevig2006c}%
  \BibitemOpen
  \bibfield  {author} {\bibinfo {author} {\bibfnamefont {B.~A.}\ \bibnamefont
  {Bernevig}}, \bibinfo {author} {\bibfnamefont {T.~L.}\ \bibnamefont
  {Hughes}}, \ and\ \bibinfo {author} {\bibfnamefont {S.-C.}\ \bibnamefont
  {Zhang}},\ }\href@noop {} {\bibfield  {journal} {\bibinfo  {journal}
  {Science}\ }\textbf {\bibinfo {volume} {314}},\ \bibinfo {pages} {1757}
  (\bibinfo {year} {2006})}\BibitemShut {NoStop}%
\bibitem [{\citenamefont {K{\"o}nig}\ \emph {et~al.}(2008)\citenamefont
  {K{\"o}nig}, \citenamefont {Buhmann}, \citenamefont {W.~Molenkamp},
  \citenamefont {Hughes}, \citenamefont {Liu}, \citenamefont {Qi},\ and\
  \citenamefont {Zhang}}]{konig2008}%
  \BibitemOpen
  \bibfield  {author} {\bibinfo {author} {\bibfnamefont {M.}~\bibnamefont
  {K{\"o}nig}}, \bibinfo {author} {\bibfnamefont {H.}~\bibnamefont {Buhmann}},
  \bibinfo {author} {\bibfnamefont {L.}~\bibnamefont {W.~Molenkamp}}, \bibinfo
  {author} {\bibfnamefont {T.}~\bibnamefont {Hughes}}, \bibinfo {author}
  {\bibfnamefont {C.-X.}\ \bibnamefont {Liu}}, \bibinfo {author} {\bibfnamefont
  {X.-L.}\ \bibnamefont {Qi}}, \ and\ \bibinfo {author} {\bibfnamefont {S.-C.}\
  \bibnamefont {Zhang}},\ }\href@noop {} {\bibfield  {journal} {\bibinfo
  {journal} {Journal of the Physical Society of Japan}\ }\textbf {\bibinfo
  {volume} {77}} (\bibinfo {year} {2008})}\BibitemShut {NoStop}%
\bibitem [{\citenamefont {Chang}\ \emph {et~al.}(2013)\citenamefont {Chang},
  \citenamefont {Zhang}, \citenamefont {Feng}, \citenamefont {Shen},
  \citenamefont {Zhang}, \citenamefont {Guo}, \citenamefont {Li}, \citenamefont
  {Ou}, \citenamefont {Wei}, \citenamefont {Wang} \emph {et~al.}}]{chang2013}%
  \BibitemOpen
  \bibfield  {author} {\bibinfo {author} {\bibfnamefont {C.-Z.}\ \bibnamefont
  {Chang}}, \bibinfo {author} {\bibfnamefont {J.}~\bibnamefont {Zhang}},
  \bibinfo {author} {\bibfnamefont {X.}~\bibnamefont {Feng}}, \bibinfo {author}
  {\bibfnamefont {J.}~\bibnamefont {Shen}}, \bibinfo {author} {\bibfnamefont
  {Z.}~\bibnamefont {Zhang}}, \bibinfo {author} {\bibfnamefont
  {M.}~\bibnamefont {Guo}}, \bibinfo {author} {\bibfnamefont {K.}~\bibnamefont
  {Li}}, \bibinfo {author} {\bibfnamefont {Y.}~\bibnamefont {Ou}}, \bibinfo
  {author} {\bibfnamefont {P.}~\bibnamefont {Wei}}, \bibinfo {author}
  {\bibfnamefont {L.-L.}\ \bibnamefont {Wang}},  \emph {et~al.},\ }\href@noop
  {} {\bibfield  {journal} {\bibinfo  {journal} {Science}\ }\textbf {\bibinfo
  {volume} {340}},\ \bibinfo {pages} {167} (\bibinfo {year}
  {2013})}\BibitemShut {NoStop}%
\bibitem [{\citenamefont {Xu}\ \emph {et~al.}(2012)\citenamefont {Xu},
  \citenamefont {Liu}, \citenamefont {Alidoust}, \citenamefont {Neupane},
  \citenamefont {Qian}, \citenamefont {Belopolski}, \citenamefont {Denlinger},
  \citenamefont {Wang}, \citenamefont {Lin}, \citenamefont {Wray} \emph
  {et~al.}}]{xu2012observation}%
  \BibitemOpen
  \bibfield  {author} {\bibinfo {author} {\bibfnamefont {S.-Y.}\ \bibnamefont
  {Xu}}, \bibinfo {author} {\bibfnamefont {C.}~\bibnamefont {Liu}}, \bibinfo
  {author} {\bibfnamefont {N.}~\bibnamefont {Alidoust}}, \bibinfo {author}
  {\bibfnamefont {M.}~\bibnamefont {Neupane}}, \bibinfo {author} {\bibfnamefont
  {D.}~\bibnamefont {Qian}}, \bibinfo {author} {\bibfnamefont {I.}~\bibnamefont
  {Belopolski}}, \bibinfo {author} {\bibfnamefont {J.}~\bibnamefont
  {Denlinger}}, \bibinfo {author} {\bibfnamefont {Y.}~\bibnamefont {Wang}},
  \bibinfo {author} {\bibfnamefont {H.}~\bibnamefont {Lin}}, \bibinfo {author}
  {\bibfnamefont {L.}~\bibnamefont {Wray}},  \emph {et~al.},\ }\href@noop {}
  {\bibfield  {journal} {\bibinfo  {journal} {Nat. Comm.}\ }\textbf {\bibinfo
  {volume} {3}},\ \bibinfo {pages} {1192} (\bibinfo {year} {2012})}\BibitemShut
  {NoStop}%
\bibitem [{\citenamefont {Tanaka}\ \emph {et~al.}(2012)\citenamefont {Tanaka},
  \citenamefont {Ren}, \citenamefont {Sato}, \citenamefont {Nakayama},
  \citenamefont {Souma}, \citenamefont {Takahashi}, \citenamefont {Segawa},\
  and\ \citenamefont {Ando}}]{tanakatci}%
  \BibitemOpen
  \bibfield  {author} {\bibinfo {author} {\bibfnamefont {Y.}~\bibnamefont
  {Tanaka}}, \bibinfo {author} {\bibfnamefont {Z.}~\bibnamefont {Ren}},
  \bibinfo {author} {\bibfnamefont {T.}~\bibnamefont {Sato}}, \bibinfo {author}
  {\bibfnamefont {K.}~\bibnamefont {Nakayama}}, \bibinfo {author}
  {\bibfnamefont {S.}~\bibnamefont {Souma}}, \bibinfo {author} {\bibfnamefont
  {T.}~\bibnamefont {Takahashi}}, \bibinfo {author} {\bibfnamefont
  {K.}~\bibnamefont {Segawa}}, \ and\ \bibinfo {author} {\bibfnamefont
  {Y.}~\bibnamefont {Ando}},\ }\href@noop {} {\bibfield  {journal} {\bibinfo
  {journal} {Nat. Phys.}\ }\textbf {\bibinfo {volume} {8}},\ \bibinfo {pages}
  {800} (\bibinfo {year} {2012})}\BibitemShut {NoStop}%
\bibitem [{\citenamefont {Castro~Neto}\ \emph {et~al.}(2009)\citenamefont
  {Castro~Neto}, \citenamefont {Guinea}, \citenamefont {Peres}, \citenamefont
  {Novoselov},\ and\ \citenamefont {Geim}}]{graphenereview}%
  \BibitemOpen
  \bibfield  {author} {\bibinfo {author} {\bibfnamefont {A.~H.}\ \bibnamefont
  {Castro~Neto}}, \bibinfo {author} {\bibfnamefont {F.}~\bibnamefont {Guinea}},
  \bibinfo {author} {\bibfnamefont {N.~M.~R.}\ \bibnamefont {Peres}}, \bibinfo
  {author} {\bibfnamefont {K.~S.}\ \bibnamefont {Novoselov}}, \ and\ \bibinfo
  {author} {\bibfnamefont {A.~K.}\ \bibnamefont {Geim}},\ }\href@noop {}
  {\bibfield  {journal} {\bibinfo  {journal} {Rev. Mod. Phys.}\ }\textbf
  {\bibinfo {volume} {81}},\ \bibinfo {pages} {109} (\bibinfo {year}
  {2009})}\BibitemShut {NoStop}%
\bibitem [{\citenamefont {Wan}\ \emph {et~al.}(2011)\citenamefont {Wan},
  \citenamefont {Turner}, \citenamefont {Vishwanath},\ and\ \citenamefont
  {Savrasov}}]{wan2011}%
  \BibitemOpen
  \bibfield  {author} {\bibinfo {author} {\bibfnamefont {X.}~\bibnamefont
  {Wan}}, \bibinfo {author} {\bibfnamefont {A.~M.}\ \bibnamefont {Turner}},
  \bibinfo {author} {\bibfnamefont {A.}~\bibnamefont {Vishwanath}}, \ and\
  \bibinfo {author} {\bibfnamefont {S.~Y.}\ \bibnamefont {Savrasov}},\
  }\href@noop {} {\bibfield  {journal} {\bibinfo  {journal} {Phys. Rev. B}\
  }\textbf {\bibinfo {volume} {83}},\ \bibinfo {pages} {205101} (\bibinfo
  {year} {2011})}\BibitemShut {NoStop}%
\bibitem [{\citenamefont {Hal{\'a}sz}\ and\ \citenamefont
  {Balents}(2012)}]{balentshalasz}%
  \BibitemOpen
  \bibfield  {author} {\bibinfo {author} {\bibfnamefont {G.~B.}\ \bibnamefont
  {Hal{\'a}sz}}\ and\ \bibinfo {author} {\bibfnamefont {L.}~\bibnamefont
  {Balents}},\ }\href@noop {} {\bibfield  {journal} {\bibinfo  {journal} {Phys.
  Rev. B}\ }\textbf {\bibinfo {volume} {85}},\ \bibinfo {pages} {035103}
  (\bibinfo {year} {2012})}\BibitemShut {NoStop}%
\bibitem [{\citenamefont {Young}\ \emph {et~al.}(2012)\citenamefont {Young},
  \citenamefont {Zaheer}, \citenamefont {Teo}, \citenamefont {Kane},
  \citenamefont {Mele},\ and\ \citenamefont {Rappe}}]{youngteo2012}%
  \BibitemOpen
  \bibfield  {author} {\bibinfo {author} {\bibfnamefont {S.~M.}\ \bibnamefont
  {Young}}, \bibinfo {author} {\bibfnamefont {S.}~\bibnamefont {Zaheer}},
  \bibinfo {author} {\bibfnamefont {J.~C.~Y.}\ \bibnamefont {Teo}}, \bibinfo
  {author} {\bibfnamefont {C.~L.}\ \bibnamefont {Kane}}, \bibinfo {author}
  {\bibfnamefont {E.~J.}\ \bibnamefont {Mele}}, \ and\ \bibinfo {author}
  {\bibfnamefont {A.~M.}\ \bibnamefont {Rappe}},\ }\href {\doibase
  10.1103/PhysRevLett.108.140405} {\bibfield  {journal} {\bibinfo  {journal}
  {Phys. Rev. Lett.}\ }\textbf {\bibinfo {volume} {108}},\ \bibinfo {pages}
  {140405} (\bibinfo {year} {2012})}\BibitemShut {NoStop}%
\bibitem [{\citenamefont {Wang}\ \emph {et~al.}(2012)\citenamefont {Wang},
  \citenamefont {Sun}, \citenamefont {Chen}, \citenamefont {Franchini},
  \citenamefont {Xu}, \citenamefont {Weng}, \citenamefont {Dai},\ and\
  \citenamefont {Fang}}]{wang2012}%
  \BibitemOpen
  \bibfield  {author} {\bibinfo {author} {\bibfnamefont {Z.}~\bibnamefont
  {Wang}}, \bibinfo {author} {\bibfnamefont {Y.}~\bibnamefont {Sun}}, \bibinfo
  {author} {\bibfnamefont {X.-Q.}\ \bibnamefont {Chen}}, \bibinfo {author}
  {\bibfnamefont {C.}~\bibnamefont {Franchini}}, \bibinfo {author}
  {\bibfnamefont {G.}~\bibnamefont {Xu}}, \bibinfo {author} {\bibfnamefont
  {H.}~\bibnamefont {Weng}}, \bibinfo {author} {\bibfnamefont {X.}~\bibnamefont
  {Dai}}, \ and\ \bibinfo {author} {\bibfnamefont {Z.}~\bibnamefont {Fang}},\
  }\href@noop {} {\bibfield  {journal} {\bibinfo  {journal} {Phys. Rev. B}\
  }\textbf {\bibinfo {volume} {85}},\ \bibinfo {pages} {195320} (\bibinfo
  {year} {2012})}\BibitemShut {NoStop}%
\bibitem [{\citenamefont {Liu}\ \emph {et~al.}(2013)\citenamefont {Liu},
  \citenamefont {Zhou}, \citenamefont {Wang}, \citenamefont {Weng},
  \citenamefont {Prabhakaran}, \citenamefont {Mo}, \citenamefont {Zhang},
  \citenamefont {Shen}, \citenamefont {Fang}, \citenamefont {Dai} \emph
  {et~al.}}]{liu2013}%
  \BibitemOpen
  \bibfield  {author} {\bibinfo {author} {\bibfnamefont {Z.}~\bibnamefont
  {Liu}}, \bibinfo {author} {\bibfnamefont {B.}~\bibnamefont {Zhou}}, \bibinfo
  {author} {\bibfnamefont {Z.}~\bibnamefont {Wang}}, \bibinfo {author}
  {\bibfnamefont {H.}~\bibnamefont {Weng}}, \bibinfo {author} {\bibfnamefont
  {D.}~\bibnamefont {Prabhakaran}}, \bibinfo {author} {\bibfnamefont {S.-K.}\
  \bibnamefont {Mo}}, \bibinfo {author} {\bibfnamefont {Y.}~\bibnamefont
  {Zhang}}, \bibinfo {author} {\bibfnamefont {Z.}~\bibnamefont {Shen}},
  \bibinfo {author} {\bibfnamefont {Z.}~\bibnamefont {Fang}}, \bibinfo {author}
  {\bibfnamefont {X.}~\bibnamefont {Dai}},  \emph {et~al.},\ }\href@noop {}
  {\bibfield  {journal} {\bibinfo  {journal} {arXiv preprint arXiv:1310.0391}\
  } (\bibinfo {year} {2013})}\BibitemShut {NoStop}%
\bibitem [{\citenamefont {Neupane}\ \emph {et~al.}(2013)\citenamefont
  {Neupane}, \citenamefont {Xu}, \citenamefont {Sankar}, \citenamefont
  {Alidoust}, \citenamefont {Bian}, \citenamefont {Liu}, \citenamefont
  {Belopolski}, \citenamefont {Chang}, \citenamefont {Jeng}, \citenamefont
  {Lin} \emph {et~al.}}]{neupane2013}%
  \BibitemOpen
  \bibfield  {author} {\bibinfo {author} {\bibfnamefont {M.}~\bibnamefont
  {Neupane}}, \bibinfo {author} {\bibfnamefont {S.}~\bibnamefont {Xu}},
  \bibinfo {author} {\bibfnamefont {R.}~\bibnamefont {Sankar}}, \bibinfo
  {author} {\bibfnamefont {N.}~\bibnamefont {Alidoust}}, \bibinfo {author}
  {\bibfnamefont {G.}~\bibnamefont {Bian}}, \bibinfo {author} {\bibfnamefont
  {C.}~\bibnamefont {Liu}}, \bibinfo {author} {\bibfnamefont {I.}~\bibnamefont
  {Belopolski}}, \bibinfo {author} {\bibfnamefont {T.-R.}\ \bibnamefont
  {Chang}}, \bibinfo {author} {\bibfnamefont {H.-T.}\ \bibnamefont {Jeng}},
  \bibinfo {author} {\bibfnamefont {H.}~\bibnamefont {Lin}},  \emph {et~al.},\
  }\href@noop {} {\bibfield  {journal} {\bibinfo  {journal} {arXiv preprint
  arXiv:1309.7892}\ } (\bibinfo {year} {2013})}\BibitemShut {NoStop}%
\bibitem [{\citenamefont {Wang}\ \emph {et~al.}(2013)\citenamefont {Wang},
  \citenamefont {Weng}, \citenamefont {Wu}, \citenamefont {Dai},\ and\
  \citenamefont {Fang}}]{wang2013}%
  \BibitemOpen
  \bibfield  {author} {\bibinfo {author} {\bibfnamefont {Z.}~\bibnamefont
  {Wang}}, \bibinfo {author} {\bibfnamefont {H.}~\bibnamefont {Weng}}, \bibinfo
  {author} {\bibfnamefont {Q.}~\bibnamefont {Wu}}, \bibinfo {author}
  {\bibfnamefont {X.}~\bibnamefont {Dai}}, \ and\ \bibinfo {author}
  {\bibfnamefont {Z.}~\bibnamefont {Fang}},\ }\href@noop {} {\bibfield
  {journal} {\bibinfo  {journal} {Phys. Rev. B}\ }\textbf {\bibinfo {volume}
  {88}},\ \bibinfo {pages} {125427} (\bibinfo {year} {2013})}\BibitemShut
  {NoStop}%
\bibitem [{\citenamefont {Yang}\ and\ \citenamefont
  {Nagaosa}(2014)}]{yang2014}%
  \BibitemOpen
  \bibfield  {author} {\bibinfo {author} {\bibfnamefont {B.-J.}\ \bibnamefont
  {Yang}}\ and\ \bibinfo {author} {\bibfnamefont {N.}~\bibnamefont {Nagaosa}},\
  }\href@noop {} {\bibfield  {journal} {\bibinfo  {journal} {arXiv preprint
  arXiv:1404.0754}\ } (\bibinfo {year} {2014})}\BibitemShut {NoStop}%
\bibitem [{\citenamefont {Ramamurthy}\ and\ \citenamefont
  {Hughes}(2015)}]{TSMresponse}%
  \BibitemOpen
  \bibfield  {author} {\bibinfo {author} {\bibfnamefont {S.~T.}\ \bibnamefont
  {Ramamurthy}}\ and\ \bibinfo {author} {\bibfnamefont {T.~L.}\ \bibnamefont
  {Hughes}},\ }\href {\doibase 10.1103/PhysRevB.92.085105} {\bibfield
  {journal} {\bibinfo  {journal} {Phys. Rev. B}\ }\textbf {\bibinfo {volume}
  {92}},\ \bibinfo {pages} {085105} (\bibinfo {year} {2015})}\BibitemShut
  {NoStop}%
\bibitem [{\citenamefont {Nielsen}\ and\ \citenamefont
  {Ninomiya}(1981)}]{nielsen1981}%
  \BibitemOpen
  \bibfield  {author} {\bibinfo {author} {\bibfnamefont {H.~B.}\ \bibnamefont
  {Nielsen}}\ and\ \bibinfo {author} {\bibfnamefont {M.}~\bibnamefont
  {Ninomiya}},\ }\href@noop {} {\bibfield  {journal} {\bibinfo  {journal}
  {Physics Letters B}\ }\textbf {\bibinfo {volume} {105}},\ \bibinfo {pages}
  {219} (\bibinfo {year} {1981})}\BibitemShut {NoStop}%
\bibitem [{\citenamefont {Shi}\ \emph {et~al.}(2007)\citenamefont {Shi},
  \citenamefont {Vignale}, \citenamefont {Xiao},\ and\ \citenamefont
  {Niu}}]{niu2007}%
  \BibitemOpen
  \bibfield  {author} {\bibinfo {author} {\bibfnamefont {J.}~\bibnamefont
  {Shi}}, \bibinfo {author} {\bibfnamefont {G.}~\bibnamefont {Vignale}},
  \bibinfo {author} {\bibfnamefont {D.}~\bibnamefont {Xiao}}, \ and\ \bibinfo
  {author} {\bibfnamefont {Q.}~\bibnamefont {Niu}},\ }\href@noop {} {\bibfield
  {journal} {\bibinfo  {journal} {Physical review letters}\ }\textbf {\bibinfo
  {volume} {99}},\ \bibinfo {pages} {197202} (\bibinfo {year}
  {2007})}\BibitemShut {NoStop}%
\bibitem [{\citenamefont {Xiao}\ \emph {et~al.}(2007)\citenamefont {Xiao},
  \citenamefont {Yao},\ and\ \citenamefont {Niu}}]{niuvalley}%
  \BibitemOpen
  \bibfield  {author} {\bibinfo {author} {\bibfnamefont {D.}~\bibnamefont
  {Xiao}}, \bibinfo {author} {\bibfnamefont {W.}~\bibnamefont {Yao}}, \ and\
  \bibinfo {author} {\bibfnamefont {Q.}~\bibnamefont {Niu}},\ }\href@noop {}
  {\bibfield  {journal} {\bibinfo  {journal} {Phys. Rev. Lett.}\ }\textbf
  {\bibinfo {volume} {99}},\ \bibinfo {pages} {236809} (\bibinfo {year}
  {2007})}\BibitemShut {NoStop}%
\bibitem [{\citenamefont {Zyuzin}\ and\ \citenamefont
  {Burkov}(2012)}]{zyuzin2012}%
  \BibitemOpen
  \bibfield  {author} {\bibinfo {author} {\bibfnamefont {A.}~\bibnamefont
  {Zyuzin}}\ and\ \bibinfo {author} {\bibfnamefont {A.}~\bibnamefont
  {Burkov}},\ }\href@noop {} {\bibfield  {journal} {\bibinfo  {journal} {Phys.
  Rev. B}\ }\textbf {\bibinfo {volume} {86}},\ \bibinfo {pages} {115133}
  (\bibinfo {year} {2012})}\BibitemShut {NoStop}%
\bibitem [{\citenamefont {Vazifeh}\ and\ \citenamefont
  {Franz}(2013)}]{vazifeh2013}%
  \BibitemOpen
  \bibfield  {author} {\bibinfo {author} {\bibfnamefont {M.}~\bibnamefont
  {Vazifeh}}\ and\ \bibinfo {author} {\bibfnamefont {M.}~\bibnamefont
  {Franz}},\ }\href@noop {} {\bibfield  {journal} {\bibinfo  {journal} {Phys.
  Rev. Lett.}\ }\textbf {\bibinfo {volume} {111}},\ \bibinfo {pages} {027201}
  (\bibinfo {year} {2013})}\BibitemShut {NoStop}%
\bibitem [{\citenamefont {Chen}\ \emph {et~al.}(2013)\citenamefont {Chen},
  \citenamefont {Wu},\ and\ \citenamefont {Burkov}}]{chen2013}%
  \BibitemOpen
  \bibfield  {author} {\bibinfo {author} {\bibfnamefont {Y.}~\bibnamefont
  {Chen}}, \bibinfo {author} {\bibfnamefont {S.}~\bibnamefont {Wu}}, \ and\
  \bibinfo {author} {\bibfnamefont {A.}~\bibnamefont {Burkov}},\ }\href@noop {}
  {\bibfield  {journal} {\bibinfo  {journal} {Phys. Rev. B}\ }\textbf {\bibinfo
  {volume} {88}},\ \bibinfo {pages} {125105} (\bibinfo {year}
  {2013})}\BibitemShut {NoStop}%
\bibitem [{\citenamefont {Haldane}(2014)}]{haldane2014}%
  \BibitemOpen
  \bibfield  {author} {\bibinfo {author} {\bibfnamefont {F.}~\bibnamefont
  {Haldane}},\ }\href@noop {} {\bibfield  {journal} {\bibinfo  {journal} {arXiv
  preprint arXiv:1401.0529}\ } (\bibinfo {year} {2014})}\BibitemShut {NoStop}%
\bibitem [{\citenamefont {Ran}(2010)}]{ran2010}%
  \BibitemOpen
  \bibfield  {author} {\bibinfo {author} {\bibfnamefont {Y.}~\bibnamefont
  {Ran}},\ }\href@noop {} {\bibfield  {journal} {\bibinfo  {journal} {arXiv
  preprint arXiv:1006.5454}\ } (\bibinfo {year} {2010})}\BibitemShut {NoStop}%
\bibitem [{\citenamefont {Hughes}\ \emph
  {et~al.}(2013{\natexlab{b}})\citenamefont {Hughes}, \citenamefont {Yao},\
  and\ \citenamefont {Qi}}]{hughesyaoqi}%
  \BibitemOpen
  \bibfield  {author} {\bibinfo {author} {\bibfnamefont {T.~L.}\ \bibnamefont
  {Hughes}}, \bibinfo {author} {\bibfnamefont {H.}~\bibnamefont {Yao}}, \ and\
  \bibinfo {author} {\bibfnamefont {X.-L.}\ \bibnamefont {Qi}},\ }\href@noop {}
  {\bibfield  {journal} {\bibinfo  {journal} {arXiv preprint arXiv:1303.1539}\
  } (\bibinfo {year} {2013}{\natexlab{b}})}\BibitemShut {NoStop}%
\bibitem [{\citenamefont {Pershoguba}\ and\ \citenamefont
  {Yakovenko}(2012)}]{yakovenko}%
  \BibitemOpen
  \bibfield  {author} {\bibinfo {author} {\bibfnamefont {S.~S.}\ \bibnamefont
  {Pershoguba}}\ and\ \bibinfo {author} {\bibfnamefont {V.~M.}\ \bibnamefont
  {Yakovenko}},\ }\href {\doibase 10.1103/PhysRevB.86.075304} {\bibfield
  {journal} {\bibinfo  {journal} {Phys. Rev. B}\ }\textbf {\bibinfo {volume}
  {86}},\ \bibinfo {pages} {075304} (\bibinfo {year} {2012})}\BibitemShut
  {NoStop}%
\bibitem [{\citenamefont {Chang}\ \emph {et~al.}(2014)\citenamefont {Chang},
  \citenamefont {Matsuura}, \citenamefont {Schnyder},\ and\ \citenamefont
  {Ryu}}]{chang2014}%
  \BibitemOpen
  \bibfield  {author} {\bibinfo {author} {\bibfnamefont {P.-Y.}\ \bibnamefont
  {Chang}}, \bibinfo {author} {\bibfnamefont {S.}~\bibnamefont {Matsuura}},
  \bibinfo {author} {\bibfnamefont {A.~P.}\ \bibnamefont {Schnyder}}, \ and\
  \bibinfo {author} {\bibfnamefont {S.}~\bibnamefont {Ryu}},\ }\href@noop {}
  {\bibfield  {journal} {\bibinfo  {journal} {arXiv preprint arXiv:1406.0232}\
  } (\bibinfo {year} {2014})}\BibitemShut {NoStop}%
\bibitem [{\citenamefont {Matsuura}\ \emph {et~al.}(2013)\citenamefont
  {Matsuura}, \citenamefont {Chang}, \citenamefont {Schnyder},\ and\
  \citenamefont {Ryu}}]{matsuura2013}%
  \BibitemOpen
  \bibfield  {author} {\bibinfo {author} {\bibfnamefont {S.}~\bibnamefont
  {Matsuura}}, \bibinfo {author} {\bibfnamefont {P.-Y.}\ \bibnamefont {Chang}},
  \bibinfo {author} {\bibfnamefont {A.~P.}\ \bibnamefont {Schnyder}}, \ and\
  \bibinfo {author} {\bibfnamefont {S.}~\bibnamefont {Ryu}},\ }\href@noop {}
  {\bibfield  {journal} {\bibinfo  {journal} {New Journal of Physics}\ }\textbf
  {\bibinfo {volume} {15}},\ \bibinfo {pages} {065001} (\bibinfo {year}
  {2013})}\BibitemShut {NoStop}%
\bibitem [{\citenamefont {Burkov}\ \emph {et~al.}(2011)\citenamefont {Burkov},
  \citenamefont {Hook},\ and\ \citenamefont {Balents}}]{burkov2011}%
  \BibitemOpen
  \bibfield  {author} {\bibinfo {author} {\bibfnamefont {A.~A.}\ \bibnamefont
  {Burkov}}, \bibinfo {author} {\bibfnamefont {M.~D.}\ \bibnamefont {Hook}}, \
  and\ \bibinfo {author} {\bibfnamefont {L.}~\bibnamefont {Balents}},\ }\href
  {\doibase 10.1103/PhysRevB.84.235126} {\bibfield  {journal} {\bibinfo
  {journal} {Phys. Rev. B}\ }\textbf {\bibinfo {volume} {84}},\ \bibinfo
  {pages} {235126} (\bibinfo {year} {2011})}\BibitemShut {NoStop}%
\bibitem [{\citenamefont {Phillips}\ and\ \citenamefont
  {Aji}(2014)}]{phillips2014}%
  \BibitemOpen
  \bibfield  {author} {\bibinfo {author} {\bibfnamefont {M.}~\bibnamefont
  {Phillips}}\ and\ \bibinfo {author} {\bibfnamefont {V.}~\bibnamefont {Aji}},\
  }\href@noop {} {\bibfield  {journal} {\bibinfo  {journal} {Physical Review
  B}\ }\textbf {\bibinfo {volume} {90}},\ \bibinfo {pages} {115111} (\bibinfo
  {year} {2014})}\BibitemShut {NoStop}%
\bibitem [{\citenamefont {Mullen}\ \emph {et~al.}(2014)\citenamefont {Mullen},
  \citenamefont {Uchoa},\ and\ \citenamefont {Glatzhofer}}]{uchoa2014}%
  \BibitemOpen
  \bibfield  {author} {\bibinfo {author} {\bibfnamefont {K.}~\bibnamefont
  {Mullen}}, \bibinfo {author} {\bibfnamefont {B.}~\bibnamefont {Uchoa}}, \
  and\ \bibinfo {author} {\bibfnamefont {D.~T.}\ \bibnamefont {Glatzhofer}},\
  }\href@noop {} {\bibfield  {journal} {\bibinfo  {journal} {arXiv preprint
  arXiv:1408.5522}\ } (\bibinfo {year} {2014})}\BibitemShut {NoStop}%
\bibitem [{\citenamefont {Weng}\ \emph {et~al.}(2014)\citenamefont {Weng},
  \citenamefont {Liang}, \citenamefont {Xu}, \citenamefont {Rui}, \citenamefont
  {Fang}, \citenamefont {Dai},\ and\ \citenamefont {Kawazoe}}]{weng2014}%
  \BibitemOpen
  \bibfield  {author} {\bibinfo {author} {\bibfnamefont {H.}~\bibnamefont
  {Weng}}, \bibinfo {author} {\bibfnamefont {Y.}~\bibnamefont {Liang}},
  \bibinfo {author} {\bibfnamefont {Q.}~\bibnamefont {Xu}}, \bibinfo {author}
  {\bibfnamefont {Y.}~\bibnamefont {Rui}}, \bibinfo {author} {\bibfnamefont
  {Z.}~\bibnamefont {Fang}}, \bibinfo {author} {\bibfnamefont {X.}~\bibnamefont
  {Dai}}, \ and\ \bibinfo {author} {\bibfnamefont {Y.}~\bibnamefont
  {Kawazoe}},\ }\href@noop {} {\bibfield  {journal} {\bibinfo  {journal} {arXiv
  preprint arXiv:1411.2175}\ } (\bibinfo {year} {2014})}\BibitemShut {NoStop}%
\bibitem [{\citenamefont {Kim}\ \emph {et~al.}(2015)\citenamefont {Kim},
  \citenamefont {Wieder}, \citenamefont {Kane},\ and\ \citenamefont
  {Rappe}}]{Kane2015}%
  \BibitemOpen
  \bibfield  {author} {\bibinfo {author} {\bibfnamefont {Y.}~\bibnamefont
  {Kim}}, \bibinfo {author} {\bibfnamefont {B.~J.}\ \bibnamefont {Wieder}},
  \bibinfo {author} {\bibfnamefont {C.~L.}\ \bibnamefont {Kane}}, \ and\
  \bibinfo {author} {\bibfnamefont {A.~M.}\ \bibnamefont {Rappe}},\ }\href
  {\doibase 10.1103/PhysRevLett.115.036806} {\bibfield  {journal} {\bibinfo
  {journal} {Phys. Rev. Lett.}\ }\textbf {\bibinfo {volume} {115}},\ \bibinfo
  {pages} {036806} (\bibinfo {year} {2015})}\BibitemShut {NoStop}%
\bibitem [{\citenamefont {Yu}\ \emph {et~al.}(2015)\citenamefont {Yu},
  \citenamefont {Weng}, \citenamefont {Fang}, \citenamefont {Dai},\ and\
  \citenamefont {Hu}}]{rui2015}%
  \BibitemOpen
  \bibfield  {author} {\bibinfo {author} {\bibfnamefont {R.}~\bibnamefont
  {Yu}}, \bibinfo {author} {\bibfnamefont {H.}~\bibnamefont {Weng}}, \bibinfo
  {author} {\bibfnamefont {Z.}~\bibnamefont {Fang}}, \bibinfo {author}
  {\bibfnamefont {X.}~\bibnamefont {Dai}}, \ and\ \bibinfo {author}
  {\bibfnamefont {X.}~\bibnamefont {Hu}},\ }\href {\doibase
  10.1103/PhysRevLett.115.036807} {\bibfield  {journal} {\bibinfo  {journal}
  {Phys. Rev. Lett.}\ }\textbf {\bibinfo {volume} {115}},\ \bibinfo {pages}
  {036807} (\bibinfo {year} {2015})}\BibitemShut {NoStop}%
\bibitem [{\citenamefont {Zak}(1989)}]{zak1989berry}%
  \BibitemOpen
  \bibfield  {author} {\bibinfo {author} {\bibfnamefont {J.}~\bibnamefont
  {Zak}},\ }\href@noop {} {\bibfield  {journal} {\bibinfo  {journal} {Phys.
  Rev. Lett.}\ }\textbf {\bibinfo {volume} {62}},\ \bibinfo {pages} {2747}
  (\bibinfo {year} {1989})}\BibitemShut {NoStop}%
\bibitem [{\citenamefont {King-Smith}\ and\ \citenamefont
  {Vanderbilt}(1993)}]{vanderbilt1993}%
  \BibitemOpen
  \bibfield  {author} {\bibinfo {author} {\bibfnamefont {R.}~\bibnamefont
  {King-Smith}}\ and\ \bibinfo {author} {\bibfnamefont {D.}~\bibnamefont
  {Vanderbilt}},\ }\href@noop {} {\bibfield  {journal} {\bibinfo  {journal}
  {Phys. Rev. B}\ }\textbf {\bibinfo {volume} {47}},\ \bibinfo {pages} {1651}
  (\bibinfo {year} {1993})}\BibitemShut {NoStop}%
\bibitem [{\citenamefont {Qi}\ \emph {et~al.}(2008{\natexlab{b}})\citenamefont
  {Qi}, \citenamefont {Hughes},\ and\ \citenamefont {Zhang}}]{qhz2008}%
  \BibitemOpen
  \bibfield  {author} {\bibinfo {author} {\bibfnamefont {X.-L.}\ \bibnamefont
  {Qi}}, \bibinfo {author} {\bibfnamefont {T.~L.}\ \bibnamefont {Hughes}}, \
  and\ \bibinfo {author} {\bibfnamefont {S.-C.}\ \bibnamefont {Zhang}},\
  }\href@noop {} {\bibfield  {journal} {\bibinfo  {journal} {Phys. Rev. B}\
  }\textbf {\bibinfo {volume} {78}},\ \bibinfo {pages} {195424} (\bibinfo
  {year} {2008}{\natexlab{b}})}\BibitemShut {NoStop}%
\bibitem [{\citenamefont {Hughes}\ \emph
  {et~al.}(2011{\natexlab{b}})\citenamefont {Hughes}, \citenamefont {Prodan},\
  and\ \citenamefont {Bernevig}}]{hughesprodanbernevig}%
  \BibitemOpen
  \bibfield  {author} {\bibinfo {author} {\bibfnamefont {T.~L.}\ \bibnamefont
  {Hughes}}, \bibinfo {author} {\bibfnamefont {E.}~\bibnamefont {Prodan}}, \
  and\ \bibinfo {author} {\bibfnamefont {B.~A.}\ \bibnamefont {Bernevig}},\
  }\href@noop {} {\bibfield  {journal} {\bibinfo  {journal} {Physical Review
  B}\ }\textbf {\bibinfo {volume} {83}},\ \bibinfo {pages} {245132} (\bibinfo
  {year} {2011}{\natexlab{b}})}\BibitemShut {NoStop}%
\bibitem [{\citenamefont {Coh}\ and\ \citenamefont
  {Vanderbilt}(2009)}]{vanderbiltchern}%
  \BibitemOpen
  \bibfield  {author} {\bibinfo {author} {\bibfnamefont {S.}~\bibnamefont
  {Coh}}\ and\ \bibinfo {author} {\bibfnamefont {D.}~\bibnamefont
  {Vanderbilt}},\ }\href@noop {} {\bibfield  {journal} {\bibinfo  {journal}
  {Phys. Rev. Lett.}\ }\textbf {\bibinfo {volume} {102}},\ \bibinfo {pages}
  {107603} (\bibinfo {year} {2009})}\BibitemShut {NoStop}%
\bibitem [{\citenamefont {Ceresoli}\ \emph {et~al.}(2006)\citenamefont
  {Ceresoli}, \citenamefont {Thonhauser}, \citenamefont {Vanderbilt},\ and\
  \citenamefont {Resta}}]{vanderbiltmag}%
  \BibitemOpen
  \bibfield  {author} {\bibinfo {author} {\bibfnamefont {D.}~\bibnamefont
  {Ceresoli}}, \bibinfo {author} {\bibfnamefont {T.}~\bibnamefont
  {Thonhauser}}, \bibinfo {author} {\bibfnamefont {D.}~\bibnamefont
  {Vanderbilt}}, \ and\ \bibinfo {author} {\bibfnamefont {R.}~\bibnamefont
  {Resta}},\ }\href@noop {} {\bibfield  {journal} {\bibinfo  {journal} {Phys.
  Rev. B}\ }\textbf {\bibinfo {volume} {74}},\ \bibinfo {pages} {024408}
  (\bibinfo {year} {2006})}\BibitemShut {NoStop}%
\bibitem [{\citenamefont {Fang}\ \emph {et~al.}(2015)\citenamefont {Fang},
  \citenamefont {Chen}, \citenamefont {Kee},\ and\ \citenamefont
  {Fu}}]{fang2015}%
  \BibitemOpen
  \bibfield  {author} {\bibinfo {author} {\bibfnamefont {C.}~\bibnamefont
  {Fang}}, \bibinfo {author} {\bibfnamefont {Y.}~\bibnamefont {Chen}}, \bibinfo
  {author} {\bibfnamefont {H.-Y.}\ \bibnamefont {Kee}}, \ and\ \bibinfo
  {author} {\bibfnamefont {L.}~\bibnamefont {Fu}},\ }\href@noop {} {\bibfield
  {journal} {\bibinfo  {journal} {arXiv preprint arXiv:1506.03449}\ } (\bibinfo
  {year} {2015})}\BibitemShut {NoStop}%
\end{thebibliography}

%merlin.mbs apsrev4-1.bst 2010-07-25 4.21a (PWD, AO, DPC) hacked
%Control: key (0)
%Control: author (72) initials jnrlst
%Control: editor formatted (1) identically to author
%Control: production of article title (-1) disabled
%Control: page (0) single
%Control: year (1) truncated
%Control: production of eprint (0) enabled
%

\appendix
\section{Multiple FLs and the polarization}
\label{sec:rules}
When we have multiple FLs, the problem of calculating the polarization precisely is not quite as simple because the boundary charge is decided by the overlap and filling of the low energy boundary states that are enclosed by the multiple FLs. Despite this, even in the most general setting, the polarization can be written down as a signed-sum of the various projected areas enclosed by the various FLs. As described in the main text, we showed that we can perform a simple bulk calculation to determine a set of values for these signs. However, a precise surface theorem giving the bound charge associated to the polarization change at an interface is meaningful only when the occupations of the surface states are specified (similar to the complications in Refs. \onlinecite{vanderbiltchern,TSMresponse} for the polarizations in a Chern insulator or 2D Dirac semi-metal respectively).
If the boundary occupations are precisely known, then one can determine the necessary sign for each area contribution that will determine the correct surface charge. Hence, the projected areas that determine the surface charge are decided by the geometry of the FLs, but the signs multiplying each area can differ from the bulk calculation, and depend explicitly on the boundary state occupation.

\begin{figure}
\label{fig:illustration}
  \includegraphics[width=8.5cm]{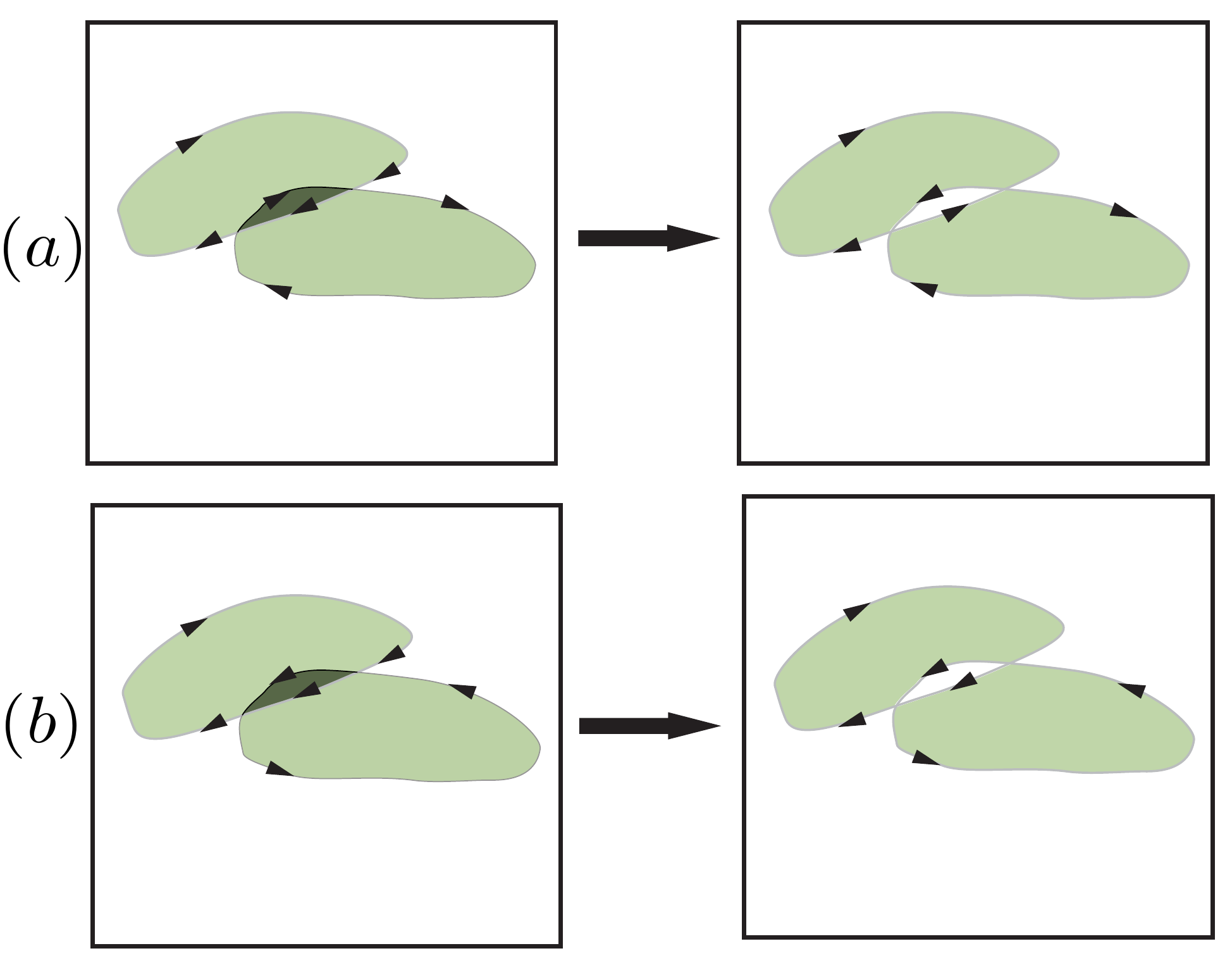}
   \caption{Rules for the modification of $\chi \mbox{sgn}\,m_{{\cal{I}}}$ for the determination of the boundary charge for the case of two FLs are illustrated. The green shaded areas represent regions where edge states exist, and the dark green area represents areas where there are overlapping edge states. Case (a) needs a reassignment of arrows while case (b) does not.}
\end{figure}

The results simplify when there is only one or two FLs in the system. In the former case, the surface charge is determined (up to a sign decided by the inversion-symmetry breaking) by whether the surface states exist inside or outside the FL. For two (or more) FLs another complication appears due to the possibility of the projected areas overlapping in the surface BZ. In these cases we can have edge states overlapping, and we expect generically that a $\mathbb{Z}_{2}$ cancelation will occur for the overlapping states. Now, let us show how we can determine the bulk value of the polarization precisely for the case of two line nodes. A natural guess for a generalization of the polarization formula we have derived in Eq.~\ref{eq:string_pol} would be 
\be
P^{i}=\epsilon^{ijk}(-1)^{\nu_{jk}}\sum_{a}\frac{e}{8\pi^{2}}\Xi_{a}\Omega_{a,jk},
\ee\noindent but this unfortunately does not account for the possible $\mathbb{Z}_{2}$ cancelations. To account for this we start off by drawing the projected FLs in the appropriate surface BZ perpendicular to the polarization direction. We must take care to include arrows indicating the direction along which Berry flux is flowing along the FL with respect to the surface normal. The flow is clockwise when the product $\chi \mbox{sgn}\,m_{{\cal{I}}}=+1$ and counterclockwise for the product $\chi \mbox{sgn}\,m_{{\cal{I}}}=-1$ where $\chi$ corresponds to the FL helicity with respect to the normal along the $i$th direction. If there are some regions where the projected areas of the FLs overlap, we have to carefully handle the $\mathbb{Z}_{2}$ cancellation. We assume that any place where two FL areas overlap there is a cancellation. We  can effectively take into account in our formula after performing a simple graphical analysis. First, if the weak invariant $(-1)^{\nu_{ij}}=-1$, we start off by shading the region around $(\pi,\pi)$, else we leave it unshaded. Then every time we cross a FL, we change from shaded to unshaded and vice versa. This prescription gives us a unique way of shading the entire surface BZ with the projected FLs where \emph{alternating} regions are shaded. The shaded regions naturally represent regions of the surface BZ with stable surface states. After we are done with shading, we check if the regions which are shaded have an arrow consistently going clockwise/counterclockwise on its boundary. If they do, we sum over the areas of the regions shaded with the product $\chi \mbox{sgn}\,m_{{\cal{I}}}$ for that region coming from the direction of the arrow on the boundary. If the direction of arrows is inconsistent, we follow the reassignment of the arrows as shown in Fig.~\ref{fig:illustration} and sum over the modified areas. 

With more FLs, this prescription does not give us a unique answer in regions which have \emph{more} than two sets of edge states overlapping. The sign of the polarization arising from these regions depends on the details of how the surface states are coupled to give the $\mathbb{Z}_{2}$ cancelation, and hence how the states are occupied. The value of the polarization that matches the surface charge is ultimately still a signed sum of the projected areas, but these signs can only be determined after the occupation of the edge state branches is chosen. All of these issues arise due to the $\mathbb{Z}_{2}$ stability of the edge states, as opposed to the $\mathbb{Z}$ stable chiral case. We will leave the problem of exhaustive treatment of generic FL configurations to future work.

\section{Magnetization in a LTSM}
\label{sec:magcalc}
Let us now calculate the magnetization for our model, which will eventually lead us to the generic form for all LTSMs. The calculation of the (orbital) magnetization in crystalline systems was developed in Refs. \onlinecite{niu2007,vanderbiltmag}, and the result of our calculation is essentially an extension of the results of the 2D Dirac semimetal shown in Refs. \onlinecite{niuvalley, TSMresponse}.
To proceed, the adiabatic (Berry) curvatures $\mathcal{F}_{xy}, \mathcal{F}_{yz}, \mathcal{F}_{zx}$ for the following generic two-band model are calculated:
\be
\label{eq:test}
H(k)=A(\vec{k})\sigma^{x}+m_{{\cal{I}}}\sigma^{y}+B(\vec{k})\sigma^{z}
\ee\noindent
where $m_{{\cal{I}}}$ represents an infinitesimal inversion-breaking mass term that must be added to properly calculate the magnetization. Note that for the purposes of calculating the adiabatic curvatures, the additional $\epsilon(\vec{k})\mathbb{I}$ term that we will add to change the energy of the FL can be ignored since its inclusion will not affect the Bloch wavefunctions. The adiabatic curvature can be represented by defining the unit vector $\hat{d}$ as
\be
\hat{d}(\vec{k})=\frac{(A,m_{{\cal{I}}},B)}{\sqrt{A^{2}+m_{{\cal{I}}}^{2}+B^{2}}}
\ee
which yields
\be
\mathcal{F}_{ij}=\epsilon^{abc}\hat{d}_{a}\del_{i}\hat{d}_{b}\del_{j}\hat{d}_{c}
\ee
where $\del_{i}=\tfrac{\del}{\del k_{i}}$ for $i=x,y,z$.
So for the model in Eq.~\ref{eq:test} we have
\be
\mathcal{F}_{ij}=m_{{\cal{I}}}\frac{\del_{i}A\,\del_{j}B-\del_{j}A\,\del_{i}B}{(A^{2}+m_{{\cal{I}}}^{2}+B^{2})^{3/2}}.
\ee
For the case of the semimetal, the limit of $m_{{\cal{I}}}\rightarrow 0$ must be taken. Using the identity that $\lim_{\epsilon\rightarrow 0}\tfrac{\epsilon}{\epsilon^{2}+\alpha^{2}}=\pi\,\mbox{sgn}\,m_{{\cal{I}}}\,\delta(\alpha)$, the curvature can be simplified to
\be
\label{eq:berrycurv}
\mathcal{F}_{ij}=\pi\,\mbox{sgn}\,m_{{\cal{I}}}\delta(\sqrt{A^{2}+B^{2}})\frac{\del_{i}A\,\del_{j}B-\del_{j}A\,\del_{i}B}{\sqrt{A^{2}+B^{2}}}.
\ee
If we think about the actual terms $A(\vec{k})$ and $B(\vec{k})$ from the model $H_3,$ then we quickly see that the $\delta$-function only has non-zero support exactly on the line-nodes. Generically, when $A(\vec{k})$ and $B(\vec{k})$ both vanish, then the system is gapless (when $m_{{\cal{I}}}\to 0$), and these gapless regions are the only sources of adiabatic curvature for a system with ${\mathcal{TI}}$ symmetry. Thus, in the gapless, semimetallic limit the only adiabatic curvature in the BZ is localized exactly on the FL, which we know must be the case for a model with $\mathcal{TI}$ symmetry.

To finish the magnetization calculation, consider the model  $\bar{H}_3(\vec{k})=\epsilon(\vec{k})\mathbb{I}+H_3(\vec{k})$ which now has broken ${\mathcal{T}}$ and broken ${\mathcal{I}},$ but preserves $\mathcal{TI}.$ The expression for the magnetization density in terms of Bloch bands is given by\cite{vanderbiltmag} 
\be
M^{a}=\epsilon^{abc}\frac{e}{2\hbar}\int \frac{d^3k}{(2\pi)^3}{\rm{Im}}\langle\partial_b u_{-}\vert( \bar{H}_3(k)+E_{-}(k))\vert\partial_c u_{-}\rangle
\ee\noindent where $E_{-}(k), \vert u_{-}\rangle$ are the energy and Bloch functions of the lower occupied band, and the derivatives are with respect to momentum. From symmetry, and from the fact that the extra kinetic term is proportional to the identity matrix, the above expression simplifies to 
\be
\label{eq:mag}
M^{a}=\,\mbox{sgn}\,m_{{\cal{I}}}\frac{e\epsilon^{abc}}{4\hbar}\int_{BZ} \frac{d^{3}k}{(2\pi)^{3}}2\epsilon(\vec{k})\mathcal{F}_{bc}.
\ee\noindent
The expression from Eq.~\ref{eq:berrycurv} for the curvature can now be substituted. Notice that we can do a coordinate transformation under the integral from $(k_{a},k_{b},k_{c})\rightarrow(k_{a},A,B)$ and the Jacobian of the transformation $J=\vert\del_{i}A\del_{j}B-\del_{j}A\del_{i}B\vert$ is already sitting in the curvature up to a total sign. Using the property that $\int_{\mathbf{X}}\delta(g(x))f(g(x))\vert g'(x)\vert dx=\int_{g(\mathbf{X})}\delta(u)f(u)du$, we can rewrite Eq.~\ref{eq:mag} as 
\be
M^{a}=\pm\,\mbox{sgn}\,m_{{\cal{I}}}\frac{e}{4\hbar}\int \frac{dk^{a}dAdB}{(2\pi)^{2}}2\epsilon(\vec{k})\frac{\delta(\sqrt{A^{2}+B^{2}})}{\sqrt{A^{2}+B^{2}}}
\ee\noindent where the domain of integration has now changed to the range of values which $A,B$ take over the BZ and the outer signs represent the helicity of the FL, i.e. the sign of the Jacobian. We can make a coordinate transformation to polar coordinates in $A,B\rightarrow r,\theta$ where we note that $r,\theta$ could in general depend on $k^{a}$. 
\be
M^{a}=\pm\,\mbox{sgn}\,m_{{\cal{I}}}\frac{e}{4\hbar}\int \frac{dk^{a}\times rdrd\theta}{(2\pi)^{2}}2\epsilon(\vec{k})\frac{\delta(r)}{r}
\ee\noindent
which can be simplified by integrating the expressions over $r,\theta$. The $\delta$ function localizes the integral to the FL and the integral over $\theta$ gives us a factor of $2\pi$. 
\be
\label{eq:string_mag}
\vec{M}=\pm\,\mbox{sgn}\,m_{{\cal{I}}}\frac{e}{4\pi\hbar}\int_{\del R}\epsilon(\vec{k})d\vec{k}
\ee\noindent
where we have explicitly indicated that the integration in Eq.~\ref{eq:string_mag} is over the FL which is equivalent to $\del R$. We note that the magnetization results from integrating the energy of each point on the FL along the line node. Again, the $\pm$ sign in front of the magnetization tells us the sense in which the Berry flux circulates along the string, i.e., clockwise or counter-clockwise. This is a simple derivation of the bulk magnetization in the case of a single line node. If there are multiple FLs, contributions to the magnetization from each FL using Eq.~\ref{eq:string_mag} must be added up, but the result is not as complicated as the polarization with multiple FLs since the magnetization adds up normally, not as a $Z_2$ quantity. It is important to note that the connection between the bulk magnetization calculation and the boundary current can depend on the details of how the boundary states are filled similar to what was shown in Refs. \onlinecite{TSMresponse} for 2D Dirac semi-metals.

\end{document}